\documentclass[a4paper, english, 11pt]{article}
\usepackage{subfiles}

\usepackage[utf8]{inputenc}
\usepackage[T1]{fontenc}
\usepackage[english]{babel}
\usepackage{csquotes}
\usepackage{textcomp}
\usepackage{varioref}
\usepackage{amsmath}
\usepackage{amssymb}
\usepackage{amsthm}
\usepackage{graphicx}
\usepackage[section]{placeins}
\usepackage{listings}
\usepackage{color}
\usepackage{titlesec}
\usepackage{tikz}
\usetikzlibrary{patterns,decorations}
\usepackage{algpseudocode}
\usepackage{algorithmicx}
\usepackage{algorithm}
\usepackage{hyperref}
\usepackage{enumitem}
\usepackage{subfig}
\usepackage{longtable}
\usepackage{caption}
\usepackage{bookmark}
\usepackage{multirow}
\usepackage{booktabs}

\definecolor{dkgreen}{rgb}{0,0.6,0}
\definecolor{gray}{rgb}{0.5,0.5,0.5}
\definecolor{mauve}{rgb}{0.58,0,0.82}

\makeatletter
\newtheorem*{rep@theorem}{\rep@title}
\newcommand{\newreptheorem}[2]{%
\newenvironment{rep#1}[1]{%
 \def\rep@title{#2 \ref{##1}}%
 \begin{rep@theorem}}%
 {\end{rep@theorem}}}
\makeatother

\theoremstyle{plain}

\newreptheorem{myprop}{Proposition}

\theoremstyle{definition}

\newtheorem*{myremark}{Remark}

\newcommand{\Unif}{\text{\normalfont{Unif}}}

\newcommand{\Multinom}{\text{\normalfont{Multinom}}}

\newcommand{\E}{\mathbb{E}}

\newcommand{\p}{\mathbb{P}}
\newcommand{\iidsim}{\overset{iid}{\sim}}

\newcommand{\bS}{\mathbf{S}}
\newcommand{\bu}{\mathbf{u}}

\newcommand{\bv}{\mathbf{v}}

\newcommand{\bx}{\mathbf{x}}

\newcommand{\bz}{\mathbf{z}}

\newcommand{\bmu}{\boldsymbol{\mu}}
\newcommand{\bsigma}{\boldsymbol{\sigma}}

\newcommand{\bSigma}{\boldsymbol{\Sigma}}
\newcommand{\bGamma}{\boldsymbol{\Gamma}}

\newcommand{\hatalpha}{\hat{\alpha}}
\newcommand{\hatlambda}{\hat\lambda}

\newcommand{\hatbmu}{\hat{\boldsymbol{\mu}}}

\newcommand{\hatbSigma}{\hat{\boldsymbol{\Sigma}}}

\newcommand{\hatbv}{\hat{\mathbf{v}}}

\newcommand{\tildebx}{\tilde{\mathbf{x}}}

\newcommand{\tildebv}{\tilde{\mathbf{v}}}

\newcommand{\diag}{\text{\normalfont{diag}}}

\newcommand{\argmax}{\text{\normalfont{argmax}}}

\makeatletter
\newcommand*{\tp}{%
{\mathpalette\@tp{}}%
}
\newcommand*{\@tp}[2]{%
\raisebox{\depth}{$\m@th#1\intercal$}%
}
\makeatother

\algnewcommand\algorithmicinput{\textbf{Input:}}
\algnewcommand\INPUT{\item[\algorithmicinput]}

\algnewcommand\algorithmicmyreturn{\textbf{Return:}}
\algnewcommand\myRETURN{\item[\algorithmicmyreturn]}

\graphicspath{{"./Fig/"}}
\usepackage{natbib}
\providecommand{\main}{.}
\usepackage{\main/JASA_manu}
\setcitestyle{authoryear,aysep={},yysep={,}}

\newcommand{\killpunct}[1]{}

\usepackage[margin=1in]{geometry}
\usepackage{setspace}
\newcommand{\blind}{0}

\begin{document}
%%%%%%%%%%%%%%%%%%%%%%%%%%%%%%%%%%%%%%%%%%%%%%%%%%%%%%%%%%%%%%%%%%%%%%%%%%%%%%
\if1\blind
{
  \bigskip
  \bigskip
  \bigskip
  \begin{center}
%   {\LARGE\bf Tailoring PCA for detecting sparse changes online in high-dimensional data streams}
    {\Large\bf Online Detection of Sparse Changes in High-Dimensional Data Streams Using Tailored Projections}
  \end{center}
  \medskip
} \fi

\if0\blind
{
  \title{\bf Online Detection of Sparse Changes in High-Dimensional Data Streams Using Tailored Projections}
%   \title{\bf Tailoring PCA for detecting sparse changes online in high-dimensional data streams}
  \author{Martin Tveten
  and Ingrid K. Glad, \\
  Department of Mathematics, University of Oslo}
  %\author{Author 1\thanks{
  %  The authors gratefully acknowledge \textit{please remember to list all relevant funding sources in the unblinded version}}\hspace{.2cm}\\
  %  Department of YYY, University of XXX\\
  %  and \\
  %  Author 2 \\
  %  Department of ZZZ, University of WWW}
  \maketitle
} \fi
%%%%%%%%%%%%%%%%%%%%%%%%%%%%%%%%%%%%%%%%%%%%%%%%%%%%%%%%%%%%%%%%%%%%%%%%%%%%%%

\begin{abstract}
% Maximum 200 words. Now: 181.
% State problem.
% Strategy of interest.
% Main take-aways.
When applying principal component analysis (PCA) for dimension reduction, the most varying projections are usually used in order to retain most of the information.
For the purpose of anomaly and change detection, however, the least varying projections are often the most important ones.
In this article, we present a novel method that automatically tailors the choice of projections to monitor for sparse changes in the mean and/or covariance matrix of high-dimensional data.
A subset of the least varying projections is almost always selected based on a criteria of the projection's sensitivity to changes.

Our focus is on online/sequential change detection, where the aim is to detect changes as quickly as possible, while controlling false alarms at a specified level.
A combination of tailored PCA and a generalized log-likelihood monitoring procedure displays high efficiency in detecting even very sparse changes in the mean, variance and correlation.
We demonstrate on real data that tailored PCA monitoring is efficient for sparse change detection also when the data streams are highly auto-correlated and non-normal.
Notably, error control is achieved without a large validation set, which is needed in most existing methods.
\end{abstract}
\textbf{Keywords}: Statistical Process Control (SPC), Principal Component Analysis, Anomaly Detection, Change-point Detection, Bootstrap/Resampling. \\
\textbf{R packages}: \texttt{tpca}, \texttt{tpcaMonitoring} and \texttt{tdpcaTEP} are available from https://github.com/Tveten. The packages include all code to easily reproduce our results. \\
\\
Additional supplementary materials are available online (see the list at the end of the article).

\newpage

% \tableofcontents
% \newpage

\section{Introduction}
\subsection{Motivation}
The exploding availability of cheap sensors has created a need for new methods to harvest insight from them.
In many applications, these sensors are deployed in large networks for online monitoring of a system.
Concrete examples include temperature monitoring of a data center at Johns Hopkins \citep{mishin_real_2014},
plant-wide monitoring of industrial processes \citep{ge_review_2017}
and semiconductor manufacturing \citep{zou_efficient_2014}.
Similar technology is also used within video segmentation \citep{kuncheva_pca_2014}, solar flare detection \citep{liu_adaptive_2015},
medical monitoring, DNA protein sequence analysis, network intrusion detection and speech recognition.
Our own motivation comes from condition monitoring of ships, where around 100-500 sensors are placed to measure the ship's state in terms of propulsion, temperatures, pressure and other physical quantities.

For many applications, there is a need for quick detection of anomalies that arise in the form of sustained changes in the data distribution.
E.g., the pressure in a valve is too high, which should be attended to as quickly as possible, or a small number of sensors suddenly becomes faulty.
This illustrates that changes may (and perhaps most often) only occur in a small subset of the sensors in an entire network.
Thus, lately, several authors (see Section \ref{lab:prior_work}) have worked on the problem of change detection from the angle of only a small, unknown set of affected sensors, 
or so-called \textit{sparse} changes.
Mostly, they focus on changes in the mean of independent normal data, or assume all parameters in the model to be known, both before and after a change.

However, in some applications \citep{hawkins_multivariate_2009, woodall_current_2014, kuncheva_pca_2014},
sparse changes both in the mean and in the covariance matrix of the sensors are of interest.
For instance, if a certain level of stability in a process is required, or because one has learned from historical data or experts that a group of sensors should be correlated in a specific way.
%We also hypothesize that the need for such methods is stronger than what it would seem like from the litterature because there are very few methods that tackle such general changes in distribution sequentially.
Additonally, parameters are unknown in most cases and must be estimated.
If estimation uncertainty is not accounted for, many false alarms will be raised, which is highly undesirable.
The problem we address in this article is therefore sequential detection of sparse changes in the mean and/or covariance structure of high-dimensional data streams, with all parameters unknown.
To make the method scalable, principal component analysis is incorporated and studied within this change detection framework.

\subsection{Problem Formulation} \label{sec:1problem}
%Overarching problem: Detection of sparse changes in the mean and/or covariance matrix by projecting the data onto a few principal axes. \\
%Some basic assumptions. \\
%Known parameters under $H_0$. \\
%Hard to give a basic model here, because the first section is quite general,
%while the normal assumptions comes in later. But maybe mention this.

%To compare the performance of using the raw data directly with the various uses of PCA for sequential change detection, we will concentrate on the normal model with indepenence in time.
Imagine a system being monitored by $D$ sensors at times indexed by $t$, yielding a multivariate data stream of observations $\mathbf{x}_t \in \mathbb{R}^D$.
First, there is a training period where $m$ observations $\mathbf{x}_{-m + 1}, \ldots, \mathbf{x}_0$ of the system under normal conditions are generated.
From $t \geq 1$ the data stream $\bx_t$ is monitored \textit{online} or \textit{sequentially} for a change in its joint distribution.
The change is thought of as being a consequence of an anomaly in the system.
Importantly, the anomaly might be local, and therefore only affect a small number of sensors.
The aim is to detect these anomalies as soon as possible, but false alarms should be kept at a controlled level.
%Apart from the assumption of the distribution function, no parameters are assumed to be known, and the training set can in principle be of arbitrary size, including $m = 0$.

For simplicity, our modelling assumptions are mainly as follows, but extensions to handle time-dependency and non-normality are presented and tested in Section \ref{sec:6}.
First, there is a training period where $m$ independent $N(\bmu_0, \bSigma_0)$ observations $\bx_t$ are gathered.
% During the training period, the observations $\bx_t$ are independently $N(\bmu_0, \bSigma_0)$ distributed.
As monitoring ensues, observations keep arriving from the null distribution until a change-point $\kappa \geq 0$,
after which the distribution of $\bx_t$ changes to $N(\bmu_1, \bSigma_1)$ for all $t > \kappa$.
% Both the pre- and post-change parameters are unknown and must be estimated.
A key element is the assumption that only a subset $\mathcal{D} \subseteq \{ 1, \ldots, D \}$ of the sensors are affected by a change, following the perspective of \citet{xie_sequential_2013}.
The \textit{subset of affected streams} is defined by 
\begin{equation}
  \mathcal{D} = \{d :\; \mu_{0, d} \not= \mu_{1, d} \text{ or } (\bSigma_0)_{d, *} \not= (\bSigma_1)_{d, *} \},
  \label{eq:affected_subset}
\end{equation}
where $(A)_{d, *}$ denotes the $d$-th row of a matrix.
In other words, we assume that the change in mean vector $\bmu_1 - \bmu_0$ and/or change in covariance matrix $\bSigma_1 - \bSigma_0$ has a sparsity structure.
This \textit{sparse online change-point problem} is summarized by the following sequential hypothesis test:
\begin{equation}
\begin{split}
  H_0 : \quad & \mathbf{x}_t \sim N(\bmu_0, \bSigma_0), \quad t = -m + 1, -m + 2 \ldots \\
  H_1 : \quad & \text{There is a } \kappa \geq 0 \text{ such that } \\
  & \mathbf{x}_t \sim N(\bmu_0, \bSigma_0), \quad t = -m + 1, \ldots, \kappa \\
  & \mathbf{x}_t \sim N(\bmu_1, \bSigma_1), \quad t = \kappa + 1, \kappa + 2, \ldots,
  \label{eq:OriginalHypothesis}
\end{split}
\end{equation}
where only parameters for $d \in \mathcal{D}$ change,
and $\kappa$, $\mathcal{D}$, $\bmu_0$, $\bSigma_0$, $\bmu_1$ and $\bSigma_1$ are \textit{all unknown}.
Our primary interest is the high-dimensional, sparse problem, where $D$ is high and $|\mathcal{D}|$ is relatively small.
Ultimately, we end up with a stopping rule for \eqref{eq:OriginalHypothesis} of the form
\begin{equation}
  T = \inf\{ t : \Lambda_{t} \geq b \},
\end{equation}
where $\Lambda_{t}$ is a running test statistic of all observations, including the training set.
% 
% In line with the idea of a global monitoring scheme for local changes,
% our primary interest is the high-dimensional, sparse problem, where $D$ is high and the number of affected streams $|\mathcal{D}|$ is relatively small.

Note that the assumption of independence in time is less restrictive than one may think.
Rather than thinking of $\bx_t$ as the raw observations, they can be thought of as residuals from a spatio-temporal model, learned in advance.
The monitoring procedure then raises an alarm when the spatio-temporal model does not explain the incomming data well anymore.

% In line with the idea of a global monitoring scheme for local changes,
% our primary interest is the high-dimensional, sparse problem, where $D$ is high and the number of affected streams $|\mathcal{D}|$ is relatively small.
% Our motivation comes from condition monitoring of ships, where perhaps $100-500$ sensors are stationed.
% Local changes correspond to such scenarios as a single motor being faulty, a small group of sensors being defect, or temperatures in subsystems that behave unexpectedly.
% Needless to say, detecting such cases quickly is critical to the safety of the ship.

To describe how the stopping rules $T$ are evaluated, 
let $\p^\kappa$ and $\E^\kappa$ denote probability and expectation when there is a true change-point at $\kappa$.
In particular, $\p^\infty$ and $\E^\infty$ mean probability and expectation under $H_0$.
For a chosen monitoring length $n$, we control the \textit{probability of false alarms} (PFA) at a given level,
\begin{equation}
  \p^\infty(T \leq n) \leq \alpha.
  \label{eq:PFA}
\end{equation}
(This measure of false alarms compared to the more common average run length (ARL) is discussed in Section \ref{sec:4}.)
Then, if a change actually occurs at $\kappa$, the aim is to detect it as quickly as possible, measured by the (conditional) \textit{expected detection delay} (EDD),
\begin{equation}
  \E^\kappa[T - \kappa | T > \kappa].
  \label{eq:EDD}
\end{equation}
The EDD is the expected sample size to detect a given change.
The lower it is, the better.

If one disregards the sparsity of the change, a solution to the problem \eqref{eq:OriginalHypothesis} can be obtained through
relatively straightforward generalized likelihood ratio methodolgy \citep{sullivan_change-point_2000, hawkins_multivariate_2009}.
However, these methods are not efficient in the high-dimensional and sparse change setting for several reasons.
Firstly, they do not incorporate prior information about the sparsity of a change, yielding slow detection.
Secondly, they scale poorly with $D$ in terms of detection speed.
To see this, let $t$ denote the current time, $k < t$ be a candidate change-point, so that $t - k$ is the number of observations used in estimating $\bSigma_1$.
Then $t - k > D$ for a non-degenerate maximum likelihood estimate.
This means that the most recent candidate change-point $k$ will grow further apart from the current time $t$ as $D$ grows, resulting in very slow detection.
Even if one uses regularization techniques to circumvent a singular maximum likelihood estimate, there would still be a need of an increasing amount of observations for a reliable estimate.
Thirdly, they are not scalable computationally as $D$ grows because of the burden of computing many increasingly larger covariance matrices.

Dimension reduction tools are often employed to overcome high-dimensional challenges,
but they have not been studied much in the online change-point detection context.
Therefore, our main objective in this work is to take a common and well understood dimension reduction tool, principal component analysis (PCA),
use knowledge about how it reacts to (sparse) changes in the mean and covariance matrix, and come up with an efficient way to use it for online change detection.
%\citet{kuncheva_pca_2014} does something similar in the offline setting, which has inspired this work,
%but we provide 

Our strategy for solving the change-point problem \eqref{eq:OriginalHypothesis} with tailored PCA is as follows:
\begin{enumerate}
  \item Obtain the sample principal axes $\hatbv_j$, $j = 1, \ldots, D$, from the training set $\bx_{-m + 1}, \ldots, \bx_0$.
  (The sample principal axes are the eigenvectors of the sample covariance matrix $\hatbSigma_0$.)
  \item Figure out which of the projections onto sample principal axes that are most sensitive to a given set of relevant or possible changes. Pick the $J$ most sensitive and disregard the rest.
  \item For $t > 0$, monitor the mean and variance of the projections $y_{j, t} = \hatbv_j^\tp \mathbf{x}_t$, $j = 1, \ldots, J$.
  In this way, the problem of detecting changes in the entire covariance matrix is reduced to detecting changes in marginal variances.
\end{enumerate}
This procedure is first studied within the modelling and evaluation framework described above to get some understanding under a clean setup, 
before an extension to more realistic data is proposed in Section \ref{sec:6}.

Point (2) above is the main focus of Sections \ref{sec:3}, while point (3) together with false alarm control is handled in Section \ref{sec:4}.
Empirical results from simulation studies are presented in Section \ref{sec:5}.
Lastly, in Section \ref{sec:6}, our method is extended to tackle non-normal and time-dependent data, and benchmarked on the Tennessee Eastman process.

\subsection{Main Contributions}
There are two main contributions of this work:
\begin{enumerate}
  \item A principled approach to automatically choosing which principal axes to keep for a specific change detection task, readily implemented in an R package.
  This was an open problem posed by \citet{kuncheva_pca_2014}.
  \item An online monitoring scheme that extends the scheme of \citet{xie_sequential_2013} for sparse, positive changes in the mean of independent data, 
  to detection of sparse changes in the mean and/or covariance matrix of time-dependent data.
  Our scheme is scalable, and includes all sources of estimation uncertainty when finding a threshold that meets a specified probability of false alarms, without the need of a large validation set.
\end{enumerate}

Expanding on \citet{kuncheva_pca_2014} and \citet{tveten_which_2019}, 
we find that a subset of the least varying projections tend to be selected for a wide range of change scenarios and pre-change covariance matrices.
We also conclude that monitoring the projections $y_{jt}$ offer a solution to all the discussed shortcomings of a direct approach;
quicker detection and computation can be attained because there are less parameters to estimate online, and information about change sparsity can be incorporated in our method for choosing projections.
% However, we elaborate, nuance and, sometimes, correct this picture by letting the answer depend on the pre-change covariance matrix and a selected set of relevant changes.
% An open problem \citet{kuncheva_pca_2014} pose is a way of choosing the subset of projections to retain.
% Our procedure offers an answer to this.

\subsection{Connections with Prior Work} \label{lab:prior_work}
The work in this article intersects with many fields, including anomaly and novelty detection in the machine learning world, 
statistical offline and online change/change-point detection, and statistical process control (SPC).

As the previous section suggests, the work in this article is mainly inspired by \citet{xie_sequential_2013}, \citet{kuncheva_pca_2014} and \citet{tveten_which_2019}.
We follow \citet{xie_sequential_2013} approximately in formulating the change-point problem.
The difference is that they are interested in the case where the variance is known and constant, there is no correlation between the streams,
and only positive changes in the mean are of interest.

On the other hand, \citet{kuncheva_pca_2014} motivated the use and study of PCA for our problem by arguing that the least varying projections were the most useful through a bivariate example.
\citet{tveten_which_2019} elaborate and, sometimes, correct their picture by letting the answer depend on the pre-change covariance matrix and a more comprehensive list of possible change scenarios.
In contrast to \citet{kuncheva_pca_2014}, \citet{tveten_which_2019} traces changes that occur in the distribution of $\bx_t$ through the projection, 
and see how the distribution of the projections $y_{j, t}$ changes as a result.
We build on this to develop the general method for choosing projections to monitor online for changes presented here.
% \citet{kuncheva_pca_2014} do not look at changes in the projections directly, and do not connect this to which changes in $\bx_t$ that cause them.
% Since it is changes in the distribution of $\bx_t$ that are of interest, our approach is more informative.
% Another difference is that we study the problem from the perspective of sparse changes, which, we argue, is more relevant.
% Finally, as has already been mentioned, we also give a procedure for choosing which projections to monitor.

The problem of sequential detection of sparse changes has recieved much recent interest beyond \citet{xie_sequential_2013}, which we have drawn upon in some way or another.
%The first to pose it might be said to be \citet{mei_efficient_2010}, who consider sparse changes under the assumption of a known pre- and post-change distribution.
%Later authors who fall into the same category of assumptions are 
Most of the research in this direction, however, is either concerned with changes in the mean of independent normals (or a known covariance matrix) 
\citep{zou_efficient_2014, wang_large-scale_2015, chan_optimal_2017},
or assumes that both the pre- and post-change distributions are known
\citep{mei_efficient_2010, banerjee_data-efficient_2015, fellouris_second-order_2016}.
The work of \citet{mei_scalable_2017} is interesting and relevant in that no assumptions on the distributions are made, but it is not a fully multivariate approach.

Our motivation for the choice of performance metrics comes from discussions by \citet{lai_sequential_1995} and \citet{lai_sequential_2010}.
These works also study generalized likelihood ratio approaches where parameters have to be estimated,
discuss window lengths as well as obtaining thresholds by bootstrapping.
All of which is relevant to the present article.

All of the mentioned articles fall in a tradition that was initiated by \citet{page_test_1955} and later expanded by \citet{lorden_procedures_1971} and \citet{moustakides_optimal_1986}.
The significant contribution of \citet{siegmund_sequential_1985} should also be mentioned.

There is also a connection from our work to \citet{kirch_use_2015} and \citet{dette_likelihood_2018},
who study online change-point detection within a more recently developed theoretical framework.
They consider monitoring statistics that incorporate a training set,
and control the probability of false alarms under the asymptotic scheme of the number of training samples going to infinity.
Their setup is very general, and contains much less rigid assumptions than the works we have mentioned so far,
but does not consider sparse changes explicitly.
Moreover, we control false alarms for a finite number of training samples.

Relevant litterature also exists within stochastic process control.
\citet{hawkins_multivariate_2009} and \citet{sullivan_change-point_2000} consider the same change-point problem as in this paper,
but without incorporating an assumption about the sparsity of a change.
%Useful methodology for post-change diagnosis is included in \citet{hawkins_multivariate_2009}.
\citet{chan_cumulative_2001} also study the detection of changes in the mean and/or covariance matrix by the use of projection pursuit as a dimension reduction tool, rather than PCA,
but assume the pre-change parameters to be known.
Additionally, there are plenty of control charts based on PCA (see for example the reviews \citet{weese_statistical_2015} and \citet{rato_systematic_2016}).
These are, however, not set within the online change-point detection framework of controlling the false alarm rate and measuring detection delays, and they only handle sparse changes implicitly.

Within the realm of anomaly detection in machine learning, PCA has been used in numerous ways.
The work in \citet{qahtan_pca-based_2015} is closely related to \citet{kuncheva_pca_2014},
but they use PCA in the standard way where only the most varying projections are selected.
\citet{lakhina_diagnosing_2004} and \citet{huang_-network_2007} use PCA to detect anomalies in (traffic) networks, and, like us, they find that it is the residual subspace of PCA that is most useful.
This fact is also pointed to in the extensive review of novelty detection techniques and applications in \citet{pimentel_review_2014}.
A difference from these works to us is that what they consider as anomalies are outliers in a trained model, not changes in distribution.
And, most importantly, we do not use the entire residual subspace, but rather the subspace of it that is most sensitive to a user-defined set of relevant distributional changes.
% Thus, we conjecture that our approach can provide even more efficient detection of changes, but possibly also outliers.
Other examples of PCA-based anomaly detection procedures are \citet{ferrer_multivariate_2007},  \citet{mishin_real_2014} and \citet{harrou_improved_2015}.
None of the articles mentioned in this paragraph considers the speed of detection, which is a major difference to our objective.

\section{Tailoring the Choice of Principal Axes to Change Detection} \label{sec:3}
In this section, the insights from \citet{tveten_which_2019} about the sensitivity of projections to various changes and the dependence on the pre-change covariance matrix 
are knit together into an algorithm that decides which projections to use for a given change-point problem.
Such an automatic choice of projections is what we mean by \textit{tailoring} PCA for change detection.
In the next section we test it in the online change detection setting.

What do we mean by sensitivity to changes? 
Akin to \citet{kuncheva_pca_2014} and \citet{tveten_which_2019}, we define it by a divergence between the marginal distribution of each projection before and after a change.
Here we follow \citet{tveten_which_2019}, who use the Hellinger distance.
The squared Hellinger distance between two normal distributions $p(x) = N(x | \xi_1, \sigma_1)$ and $q(x) = N(x | \xi_2, \sigma_2)$ is given by
\[
  H^2(p, q) = 1 - \sqrt{\frac{2\sigma_1\sigma_2}{\sigma_1^2 + \sigma_2^2}} \exp \left\{ -\frac{1}{4} \frac{(\xi_1 - \xi_2)^2}{\sigma_1^2 + \sigma_2^2} \right\}.
\]
A desirable feature of the Hellinger distance between two normals is that it is symmetric with respect to whether the variance increases or decreases in the sense that
a multiplicative increase of the variance by a factor $a \geq 1$ changes the distribution as much as a decrease by the factor $1/a$.
This is also a property of the generalized likelihood ratio procedure for detecting changes in the mean and/or variance we use for monitoring later.
% Secondly, due to the normal distribution, it has the property of being sensitive to outliers, making it a natural candidate for measuring sensitivity to changes as well.
There could be reasons for using other divergences, however, so in the accompanying R package, any divergence can be specified.
Our own experiments suggest that the overall conclusions will not be significantly different by using for example the KL-divergence or Bhattacharyya distance.

Formally, the definition of sensitivity to changes we use, as defined in \citet{tveten_which_2019}, is as follows.
Recall that $\bmu_0$ and $\bSigma_0$ are the pre-change mean and covariance matrix of $\bx_t$, while $\bmu_1$ and $\bSigma_1$ are the post-change parameters.
Without loss of generality, assume that $\bx_t$ is standardized with respect to the pre-change parameters, so that $\bmu_0 = \mathbf{0}$ and $\bSigma_0$ is a correlation matrix.
Next, let $\{ \lambda_j, \bv_j\}_{j = 1}^D$ be the normalized eigensystem of $\bSigma_0$, where it has been sorted such that $\lambda_1 \geq ... \geq \lambda_D$.
Then the orthogonal projections onto the pre-change principal axes are given by $y_{j, t} = \bv_j^\tp \bx_t$, for $j = 1, \ldots, D$.
Assuming $\bx_t$ is multivariate normal, $y_{j, t}$ has marginal pre- and post-change density functions
\begin{equation}
  \begin{split}
    p_j(y) &= N(y |\; \bv_j^\tp \bmu_0, \bv_j^\tp \bSigma_0 \bv_j) = N(y |\; 0, \lambda_j) \\
    q_j(y) &= N(y |\; \bv_j^\tp \bmu_1, \bv_j^\tp \bSigma_1 \bv_j),
    \label{eq:projPdf}
  \end{split}
\end{equation}
respectively.
Given a correlation matrix $\bSigma_0$, the \textit{sensitivity of the $j$'th projection to the change specified by $(\bmu_1, \bSigma_1)$} is defined as $H(p_j, q_j)$, abbreviated by $H_j$.
Importantly, note that the sensitivity as defined here is a function of the pre- and post-change parameters of the original data $\bx_t$: $\bSigma_0$, $\bmu_1$ and $\bSigma_1$.

% Formally, assume without loss of generality that the $\bx_t$'s are multivariate normal with mean $\mathbf{0}$ and correlation matrix $\bSigma_0$.
% Denote the normalized eigensystem of $\bSigma_0$ by $\{ \lambda_j, \bv_j\}_{j = 1}^D$, where it has been sorted such that $\lambda_1 \geq ... \geq \lambda_D$.
% We refer to $\bv_j$ as the $j$'th (population) pre-change principal axis of $\bx_t$ or $\bSigma_0$.
% At some point, the distribution of $\bx_t$ changes to a multiviariate normal with mean $\bmu_1$ and $\bSigma_1$.
% As observations arrive, they are projected onto the pre-change principal axes to obtain the projections $y_{j, t} = \bv_j^\tp \bx_t$.
% Let $p_j$ and $q_j$ denote the marginal pre- and post-change density functions of $y_{j, t}$, respectively.
% They are given by
% \begin{equation}
%   \begin{split}
%     p_j(y) &= N(y |\; \bv_j^\tp \bmu_0, \bv_j^\tp \bSigma_0 \bv_j) = N(y |\; 0, \lambda_j) \\
%     q_j(y) &= N(y |\; \bv_j^\tp \bmu_1, \bv_j^\tp \bSigma_1 \bv_j),
%     \label{eq:projPdf}
%   \end{split}
% \end{equation}
% Given a correlation matrix $\bSigma_0$, the \textit{sensitivity of the $j$'th projection to the change specified by $(\bmu_1, \bSigma_1)$} is defined as $H(p_j, q_j)$, abbreviated by $H_j$.
% Importantly, note that the sensitivity as defined here is a function of the pre- and post-change parameters of the original data $\bx_t$: $\bSigma_0$, $\bmu_1$ and $\bSigma_1$.
% 
Using this definition of sensitivity, \citet{tveten_which_2019} proved that for bivariate normal data, 
the least varying projection is the most sensitive if one of the means change, one of the variances increases, or the correlation changes.
If one variance decrease, then the most varying projection is the most sensitive unless the pre-change correlation is larger than $\sqrt{3}/2 \approx 0.87$,
On the other hand, when both means or both variances change, there are no clear winner among the projections.
Thus, we hypothesize that the least varying projections are particularly useful if changes have some sparsity structure, which they almost always are in the high-dimensional setting.
The general take-away, however, is that which projections are most sensitive depends on the pre-change correlation matrix and the exact nature of the change.

The tailored PCA (TPCA) method is motivated from the bivariate results.
In short, the procedure is as follows.
First, an estimate of the pre-change correlation matrix, $\hatbSigma_0$, must be obtained from a training set.
Then simulate $B$ changes from a customizable \textit{change distribution} $p(\bmu_1, \bSigma_1 | \hatbSigma_0)$, measure each projection's sensitivity to each change,
$(H_1, \ldots, H_D)^{(b)}$, $b = 1, \ldots, B$, and summarize the sensitivity in a way that yields a meaningful ranking of the principal axes/projections.
A selection of projections can then be made from the ranking.

In principle, any distribution for $p(\bmu_1, \bSigma_1 | \hatbSigma_0)$ could be used,
but the space of all possible combinations of changes is extremely vast. 
%all possible combinations of affected parameters, as well as all possible values that the parameters can take within each combination.
Therefore, we make some restrictions to simplify the space of changes.
First, we restrict ourselves to consider only one \textit{change type} at a time. %, where the distinct types are changes in the mean, variance and correlation.
The change type can then be seen as a single-trial multinomially distributed random variable $\mathbf{C}$ with probabilities $p_\mu$ (mean), $p_\sigma$ (variance) and $p_\rho$ (correlation).
Secondly, let $K = |\mathcal{D}| \in \{1, 2, \ldots, D \}$ be the \textit{change sparsity}, where $\mathcal{D}$ is defined as in \eqref{eq:affected_subset},
which indicates how many dimensions that are affected by a change.
I.e., the number of non-zero elements in $\bmu_1 - \bmu_0$ and $\bsigma_1 - \bsigma_0$, where $\bsigma$ is the diagonal of $\bSigma$. 
For a change in correlation, a change sparsity of $K$ means that all the correlations between the $K$ affected dimensions change.
Thirdly, given a change sparsity $K$, we assume throughout that the exact subset of affected streams $\mathcal{D}$ is uniformly distributed over all combinations of size $K$.
% matters since $\bSigma_0$ is not a diagonal covariance matrix.
% Therefore we assume throughout that the subset of affected streams $\mathcal{D}$ is uniformly distributed over all combinations of a given size $K$.
Fourthly, there is the \textit{change size} of each type of change:
\begin{itemize}
\item $\mu_{d} \in \mathbb{R}$ is the size of an additive change in the mean in the $d$'th component for $d \in \mathcal{D}$.
\item $\sigma_{d} \in \mathbb{R}_{>0}$ is the size of a multiplicative change in the standard deviation in the $d$'th component for $d \in \mathcal{D}$.
\item $a_{di}$ such that $a_{di}\rho_{di} \in [0, 1)$ for $d \not= i \in \mathcal{D}$ is the size of a multiplicative change in each pre-change correlation. 
  (Not all changes of this element-wise sort will result in a positive definite correlation matrix. See the supplementary material for how we deal with this.)
\end{itemize}
Note that for practical purposes it is reasonable to restrict the domain of the change sizes to sizes that are actually relevant, 
but the above outlines the theoretical scope of the post-change parameter subspace we consider.

A change distribution $p(\bmu_1, \bSigma_1 | \bSigma_0)$ can now be characterized by a distribution over the parameters ${(\mathbf{C}, K, \mathcal{D}, \mu_d, \sigma_d, a_{di})}$.
Due to space limitations, we will only show results for a change distribution that represents very little prior information about the nature of a change.
The two minor exceptions are that we assume interest is restricted to change sparsities $K \leq D/2$ and that the correlation can only decrease.
This change distribution is given by
\begin{equation}
\begin{split}
  \mathbf{C} &\sim \Multinom(p_\mu = 1/3, p_\sigma = 1/3, p_\rho = 1/3) \\
  K &\sim \Unif\{ 1, \ldots, D/2 \} \\
  \mathcal{D} | K &\sim \Unif\{\mathcal{D} \subseteq \{1, \ldots, D \} : |\mathcal{D}| = K\} \\
  \mu_d | \mathcal{D}, \mathbf{C} &\iidsim \Unif[-1.5, 1.5],\; d \in \mathcal{D} \\
  \sigma_d | \mathcal{D}, \mathbf{C} &\iidsim \frac{1}{2}\Unif[1/2.5, 1] + \frac{1}{2}\Unif[1, 2.5],\; d \in \mathcal{D} \\
  a_{di} | \mathcal{D}, \mathbf{C} &\iidsim \Unif[0, 1],\; d \not= i \in \mathcal{D}.
\end{split}
\label{eq:ex_change_distribution}
\end{equation}
The supplementary material contains simulations that show that the exhibited results are fairly robust to the choice of change distribution.
In the accompanying R-package \texttt{tpca}, one can easily set up uniform change distributions as in \ref{eq:ex_change_distribution} over other sets of change scenarios.

\begin{figure}[htb]
  \centering
  \includegraphics[scale = 0.7]{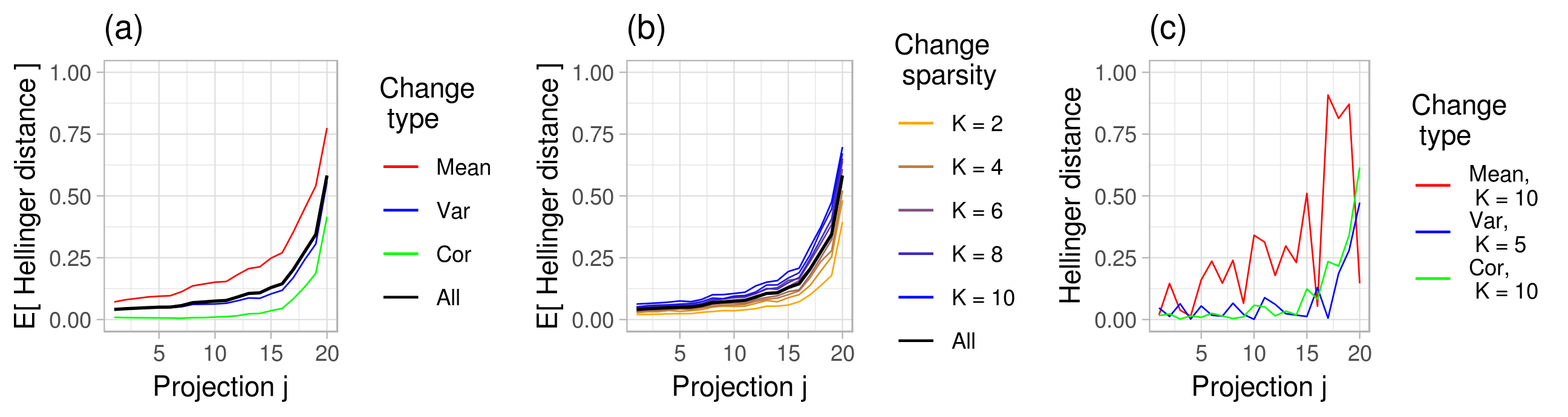}
  \caption{(a) and (b) display Monte Carlo estimates of $E[H_j|\bSigma_0]$, $j = 1, \ldots, 20$ for a randomly generated $\bSigma_0$,  
  with respect to the change distribution \eqref{eq:ex_change_distribution}.
  (a) show results conditional on change type and (b) on change sparsity.
  $10^4$ Monte Carlo samples were used.
  Note that $j = 1$ and $j = 20$ are the most and least varying projections, respectively.
  %We see that, on average, the least varying projections are more sensitive to changes than the most varying, both within each class of change types as well as for different change sparsities.
  (c) displays $H_j$ for one randomly selected change in each class of change type, to illustrate what each outcome that is averaged over to obtain (a) and (b) looks like.}
  \label{fig:ex_hellinger}
\end{figure}

% * X Define sensitivity to changes.
% * X Summarize findings in CITE[Tveten (2019)]
% * X Idea for tailoring: Draw many changes and record the most sensitive axis. Average to obtain probability estimates.
% * X Change distribution
% 
% Using change distribution \eqref{eq:ex_change_distribution} For $D = 20$ and $D = 100$, 
% 
% \begin{figure}[ht]
%   \centering
%   \includegraphics[scale = 0.7]{avg_hellinger_dense-d100}
%   \caption{Monte Carlo estimates of $E[H_j]$ with respect to the change distribution \eqref{eq:ex_change_distribution} and uniformly drawn pre-change correlation matrices $\bSigma_0$.
%   The left plot shows results conditional on change type, while the right plot displays estimates within the different change sparsities.
%   $10^3$ $\bSigma_0$'s were drawn by the method of \citet{joe_generating_2006}, and $10^3$ simulated changes were used for each $\bSigma_0$.}
%   \label{fig:avg_hellinger}
% \end{figure}
% 

By using change distribution \eqref{eq:ex_change_distribution} and a randomly generated 20-dimensional pre-change correlation matrix, 
Figure \ref{fig:ex_hellinger} illustrates that the least varying projections are the most sensitive on average,
but that notable variation is hidden on the level of the exact nature of a change.
To capture this variation, our idea is to estimate how often projection $j$ is the most sensitive one for a given correlation matrix, and use this to rank the projections.
That is, we want to estimate
\begin{equation}
  P_j := \p\left( \underset{i \in \{1, \ldots, D\}}{\argmax} H(p_i, q_i) = j \Big|\; \bSigma_0\right), \text{ for } j = 1, \ldots, D.
  \label{eq:p_argmax}
\end{equation}
%rather than $E[H_j|\bSigma_0]$.
In this way, the probability of omitting a projection that is maximally sensitive to a particular change can be controlled.
% The heuristic behind using these probabilities as a ranking is that \eqref{eq:p_argmax} records which projections are the most sensitive when the least varying axis is not.
% In this way it can be avoided with high probability that certain changes go ``under the radar''.

\begin{figure}[ht]
\centering
\includegraphics[scale = 0.7]{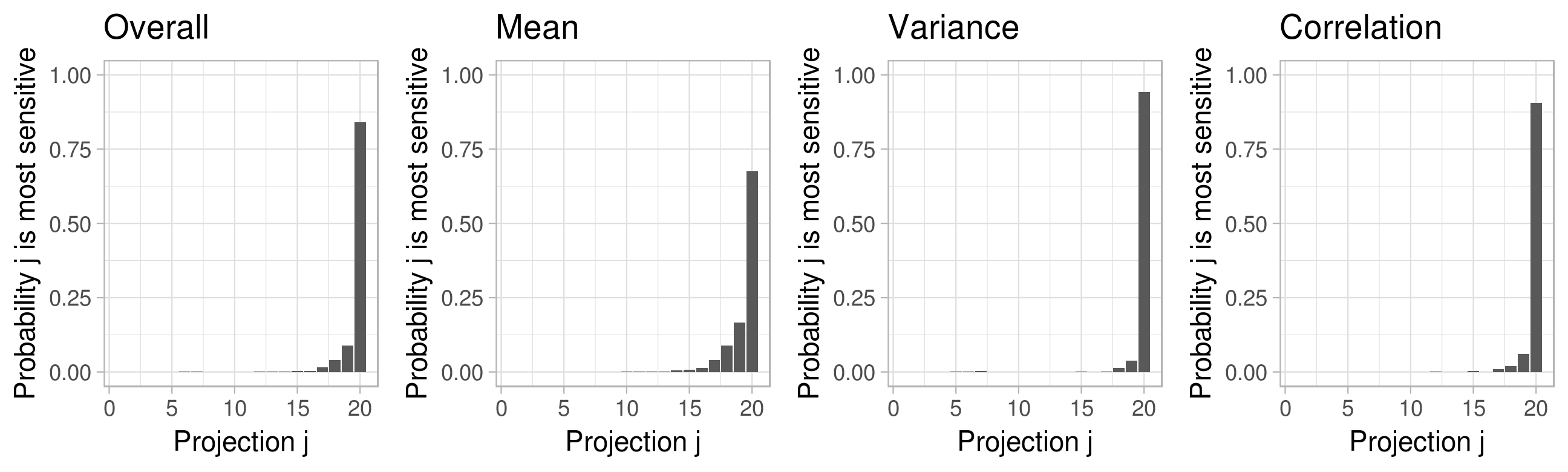}
\caption{Monte Carlo estimates of $P_j$ with respect to the same $\bSigma_0$ as in Figure \ref{fig:ex_hellinger} and same draws from change distribution \eqref{eq:ex_change_distribution}. 
The three right-most figures show the contributions to the overall probabilities (left) for each change type.}
\label{fig:p_j}
\end{figure}

To automate the choice of a projections, a cutoff value $c \in [0, 1]$ can be selected such that the projections with the highest probability of being the most sensitive are picked until the sum of probabilities is greater than $c$.
Then $1 - c$ corresponds to the probability of not picking a projection that is maximally sensitive to a change.
Figure \ref{fig:p_j} displays the estimated probabilities $\hat{P}_j$ corresponding to the same simulations as in Figure \ref{fig:ex_hellinger}.
Observe that even for $c$ close to 1, only a few of the least varying projections would be selected for all change types.
However, more axes would be selected for general changes and changes in the mean than for changes in the variance or correlation.
Consult the supplementary material for more simulations regarding which axes that are selected.

% 
% For the covariance matrix in the example of \ref{sec:2_given} and change distribution \eqref{eq:ex_change_distribution},
% the estimated probabilities $\hat{P_j}$ are shown in Figure \ref{fig:p_j}.
% Observe that the least varying projection is the most sensitive one for approximately $84\%$ of the changes.
% For a change in variance and correlation it is the most sensitive axis $94\%$ and $90\%$ of the times, respectively,
% while for the mean, a range of other projections are the most sensitive for approximately $33\%$ of the simulated changes.

% The resulting ranking of the principal axes can then be used to automate the choice of which principal axes to keep for change detection.
% A simple way is to choose the $J$ maximal probability masses of \eqref{eq:p_argmax}.
% However, perhaps a more informative way to choose the axes is to set a threshold proportion $c \in [0, 1]$,
% and choose the $J_c$ axes with maximal probability mass that sums to more than or equal to $c$.
% The interpretation of this $c$ is that $1 - c$ is the probability of not projecting the data onto an axis that is the maximally sensitive to a change.
% Thus, it quantifies the risk of reducing the dimensionality.
% This way of choosing axes is also similar in flavor to the regular use of PCA where one chooses a proportion of variance to discard.
% 
To summarize, Algorithm \ref{alg:TailoredPCA} describes the tailoring procedure in detail.
We call it the TPCA algorithm, and it is implemented in the R package \texttt{tpca}.
For online monitoring of data streams, it is intended as a final step in the training phase.
In the training phase, an estimate $\hatbSigma_0$ of the pre-change correlation matrix is obtained, 
a change distribution $p(\bmu_1, \bSigma_1 | \hatbSigma_0)$ is set up to represent the changes of interest, 
and a cutoff $c$ is chosen.
Then the tailoring algorithm is run to determine which principal axes $\mathcal{J} \in \{ 1, \ldots, D \}$ to project the incomming data onto.
Ultimately, monitoring of $\hatbv_j^\tp \bx_t$, $j \in \mathcal{J}$ ensues, which we deal with next.

% TODO: Make the tailoring point clearer.

\begin{algorithm} 
  \caption{Tailored PCA (TPCA) for Change Detection} \label{alg:TailoredPCA}
  \begin{algorithmic}[1]
    \INPUT{$\bSigma_0$, $p(\bmu_1, \bSigma_1 | \bSigma_0)$, $c$, $B$}
    \State Compute (sorted) eigenvalues and eigenvectors $\{ \lambda_j, \mathbf{v}_j \}_{j = 1}^D$ of $\bSigma_0$
    %\State Sort $\{ \lambda_j, \mathbf{v}_j \}_{j = 1}^d$ by $\lambda_1 \geq \lambda_2 \geq \ldots \geq \lambda_D$    
    %\State $\mathbf{V} \gets \begin{pmatrix} \mathbf{v}_1 & \cdots & \mathbf{v}_d \end{pmatrix}$
    \For{$b \in \{1, \ldots, B \}$}
      \State $(\bmu_1, \bSigma_1)^{(b)} \sim p(\bmu_1, \bSigma_1 | \bSigma_0)$
      %\State $\mathbf{m}^{(b)} \gets \mathbf{V}^\tp \bmu_1^{(b)}$
      %\Comment Post-change means along each PA
      %\State $\mathbf{c}^{(b)} \gets \diag\{ \mathbf{V}^\tp \bSigma_1^{(b)}\mathbf{V} \}$
      %\Comment Post-change variances along each PA 
      \State $(H_1^{(b)}, \ldots, H_D^{(b)}) \gets (H(p_1, q_1^{(b)}), \ldots, H(p_D, q_D^{(b)})$
%       \For{$j \in \{1, \ldots, D \}$}
%         \State $p_j(x) \gets N(x | 0, \lambda_j)$
%         \State $q_j^{(b)}(x) \gets N\Big(x | \bv_j \bmu_1^{(b)}, \bv_j^\tp \bSigma_1^{(b)} \bv \Big)$
%         \State $H_j^{(b)} \gets H(p_j, q_j^{(b)})$
        \Comment{$p_j$ and $q_j$ are given in \eqref{eq:projPdf}.}
%       \EndFor
      %\State $\mathbf{H}^{(b)} \gets \big(H_1^{(b)}, \ldots, H_N^{(b)}\big)$
    \EndFor
    \State $\hat{P}_j \gets \displaystyle\frac{1}{B} \displaystyle\sum_{b = 1}^{B} I\left\{\underset{i \in \{ 1, \ldots, D \}}{\argmax} H_i^{(b)} = j \right\}$
    \State $\mathcal{J} \gets \{j : \sum_{i \in \mathcal{J}} \hat{P}_i \geq c \text{ such that } |\mathcal{J}| \text{ is minimal} \}$.  
%     \State $\mathcal{J} \gets \{j : \hat{P}_j \geq \hat{P}_i \text{ for all } i \not\in \mathcal{J} \text{ and } \sum_{i \in \mathcal{J}} \hat{P}_i \geq c \}$.  
    %\State $\mathbf{A} \gets (\mathbf{v}_n :\; n \in \mathbf{n}_\text{top})^\tp$
    \myRETURN{$\mathcal{J}$ and $\{\lambda_j, \bv_j\}_{j \in \mathcal{J}}$}
  \end{algorithmic}
\end{algorithm}
  
%Given a covariance matrix under the null hypothesis, and which types of changes are of interest,
%the most important principal axes can be calculated based on the machinery above.
%If all changes are of interest, an average must be taken.
%Many different tuning parameters in this case.
%How robust is the choice of principal axes to different tuning parameters?

\section{Online Monitoring} \label{sec:4}

In this section, focus is shifted back to the sparse online change-point problem \eqref{eq:OriginalHypothesis}.
First, the monitoring statistic we will use for performance analysis is presented, before we turn to handling the uncertainty stemming from estimating the eigensystem.
We have chosen a monitoring statistic that can be set up to handle sparse changes in the mean and/or variance vectors directly in the original data $\bx_t$ as well as indirectly in the projections.
In this way, we obtain a fair benchmark for the TPCA method, both in terms of detection speed and dimension reduction capabilities.
What the monitoring statistic can not do when applied directly to the original data is to detect changes in the cross-stream correlations of $\mathbf{x}_t$.
We view this ability as an advantage of PCA-based procedures.
Whether such an ability is important or not depends on the application.

%We have chosen the monitoring statistic with two questions in mind:
%\begin{enumerate}
%  \item In terms of detection speed, how well does the tpca method compare to another method that also explicitly handles sparse changes?
%  \item How much can the dimension be reduced by without compromising on detection speed?
%\end{enumerate}

\subsection{A Mixture Procedure for Detecting Changes in the Mean and/or Variance}
%The mixture detection statistic for changes in mean and/or variance.
%\begin{itemize}
  %\item Hypotheses
  %\item Derivation
  %\item How it's applied to the reduced data.
  %\item Why do we only monitor the individual streams, and not their interaction? Way too heavy computational procedure to monitor the entire multivariate distribution. Changes in correlation will be visible on the principal axes nonetheless (an advantage of the approach).
%\end{itemize}
%
% We have chosen such a monitoring statistic because it can be set up to handle sparse changes in the mean and/or variance vectors directly in the raw data, as well as indirectly through the projections.
% In this way we obtain a fair benchmark for the tpca method, both in terms of detection speed and dimension reduction capabilities.
% What the monitoring statistic can not do when applied to the raw data is to detect changes in the cross-stream correlations of $\mathbf{x}_t$.
% We view this ability as an advantage of the PCA-based procedure.
% Whether such an ability is important or not depends on the application.

Our monitoring statistic generalizes the mixture generalized likelihood ratio (GLR) detection procedure of \citet{xie_sequential_2013} 
from only detecting positive mean shifts to detecting all changes in the mean and/or variance.
The key component in their mixture procedure is the incorporation of a prior guess about the sparsity of the change.
As before, we assume there is a training set of size $m$ with observations from the null distribution available, 
and that only an unknown proportion $p = |\mathcal{D}| / D$ of the streams are affected by a change.
The mixture procedure arises from the following hypothesis testing setup:
\begin{equation}
  \begin{split}
    H_0 : \quad & \mathbf{x}_t \sim N\big(\bmu_0,\; \diag\{ \bsigma^2_0 \}\big), \quad t = -m + 1, -m + 2 \ldots \\
    H_1 : \quad & \text{There is a } \kappa \geq 0 \text{ such that } \\
    & \mathbf{x}_t \sim N\big(\bmu_0,\; \diag\{ \bsigma^2_0 \}\big), \quad t = -m + 1, \ldots, \kappa \\
    & \mathbf{x}_t \sim N\big(\bmu_1,\; \diag\{ \bsigma^2_1 \}\big), \quad t = \kappa + 1, \kappa + 2, \ldots,
  \label{eq:multiMixHypothesis}
  \end{split}
\end{equation}
where $\mu_{0, d} \not= \mu_{1, d}$ and/or $\sigma_{0, d}^2 \not= \sigma_{1, d}^2$ only for $d \in \mathcal{D} \subseteq \{ 1, \ldots, D \}$.
The key component in the mixture procedure of \citet{xie_sequential_2013} is to substitute the unknown $p$ with a prior guess $p_0$,
which acts as the probability that each stream $n$ belongs to the class of affected streams or not.
Note that it is assumed that changes occur in the mean and/or variance simultaneously, and then persist for all $t > \kappa$.

The mixture log-likelihood ratio statistic for a change in the mean and/or variance is derived in the following.
With an assumed change-point at $\kappa = k \geq 0$ the global log-likelihood ratio is on the form
\begin{equation}
  \Lambda_{k, t}(p_0) = \sum_{d = 1}^{D} \log\left[ 1 - p_0 + p_0 \exp\left\{ \ell_{d, k, t} \right\} \right],
  \label{eq:GlobalMixLogLikRatio}
\end{equation}
where $\ell_{d, k, t}$ is the maximized likelihood ratio statistic for each stream $d$.
So with probability $1 - p_0$ all observations in stream $d$ are assumed to come from the same distribution, 
while with probability $p_0$, the distribution of a stream can be different before and after $k$.
Denote the maximum likelihood estimators for the mean and variance of each stream $d$ by
\[
  \bar{x}_{d, i, l} := \frac{1}{l - i}\sum_{j = i + 1}^{l} x_{d, j} \quad \text{and} \quad S^2_{d, i, l} := \frac{1}{l - i} \sum_{j = i + 1}^{l} (x_{d, j} - \bar{x}_{d, i, l})^2.
\]
Then standard calculations lead us to
\begin{equation}
  \ell_{d, k, t} = - \frac{m + k}{2}\log \frac{S^2_{d, -m, k}}{S^2_{d, -m, t}} - \frac{t - k}{2} \log \frac{S^2_{d, k, t}}{S^2_{d, -m, t}}.
  \label{eq:MaxLogLikRatio}
\end{equation}
See for instance \citet[p.~166]{hawkins_statistical_2005}.
Note that $\Lambda_{k, t}(p_0)$ also depends on $m$ although it is suppressed in the notation.

Ideally, a change would be declared if $\max_k \Lambda_{k, t}(p_0)$ raises above a threshold $b$.
However, a minor correction is preferable to prevent unwanted behavior, namely, that declaration of a change is much more likely for small sample sizes $t - k$.
This is so because the distribution of $\Lambda_{k, t}(p_0)$ strongly depends on the number of observations used to estimate the post-change parameters.
For example, the variance of $\Lambda_{198, 200}(p_0)$ is much larger than $\Lambda_{100, 200}(p_0)$, 
making a realization from $\Lambda_{198, 200}(p_0)$ more likely to be above $b$ than $\Lambda_{100, 200}(p_0)$.
An often used remedy is to find a Bartlett correction \citep[p.~166]{hawkins_statistical_2005}, where one finds a multiplicative correction factor $C(k, t)$, 
such that the expected value of the statistic under the null hypothesis equates to its asymptotic expected value.
The asymptotic expected value of $\Lambda_{k, t}(p_0)$ under the null hypothesis is, alas, unknown.
However, the asymptotic expected value of $2\ell_{d, k, t}$ is $4$ due to the classical result by Wilks.
Using that for a chi-square distributed $X$ with $a$ degrees of freedom, $\E[\log X] = \log 2 + \psi(a / 2)$, where $\psi$ is the digamma function, 
a correction factor for each stream $d$ is given exactly by
\begin{align*}
  \E\big[ 2\ell_{d, k, t} / C(k, t) \big] =&\; 4 \\
  2C(k, t) =& - (m + t)\log (m + t) + (m + t) \psi\left([m + t - 1]/2 \right) \\
  &+ (m + k)\log (m + k) - (m + k) \psi\left([m + k - 1]/2 \right) \\
  &+ (t - k)\log (t - k) - (t - k) \psi\left([t - k - 1]/2 \right).
\end{align*}

In total, the global corrected mixture log-likelihood ratio statistic becomes
\begin{equation}
  \Lambda^C_{k, t}(p_0) := \sum_{d = 1}^{D} \log\left[ 1 - p_0 + p_0 \exp\left\{ \ell_{d, k, t} / C(k, t) \right\} \right].
  \label{eq:GlobalCorrectedMixLogLikRatio}
\end{equation}
It further defines the stopping time that constitute the detection procedure,
\begin{equation}
  T(p_0, b) := \inf \Big\{t \geq 2: \underset{0 \leq k \leq t - 2}{\max} \Lambda^C_{k, t}(p_0) \geq b \Big\}.
  \label{eq:MixCorrectedStoppingRule}
\end{equation}
% In words, this rule says ``stop the first time the likelihood ratio at the most likely change-point raises above the threshold $b$''.
For comparing performance, we can now apply $T(p_0, b)$ to the original data $\mathbf{x}_t$ with various choices of $p_0$,
and to the projections $\mathbf{y}_t$ with $p_0 = 1$ (since we want to see how TPCA handles sparsity on its own).

In practice, we will also restrict the set of $k$'s that the maximum is taken over to a set $\mathcal{K} = \{ k : 2 \leq t - k \leq w + 1 \}$,
where $w$ is called the \textit{window size} and denotes the number of previous time-points that are considered as candidate change-points.
This is to limit memory usage and not allow the algorithm to become slower and slower indefinitely as $t$ grows.
Choices for the set $\mathcal{K}$ and the effect of the window size is discussed in \citet{lai_sequential_1995}.
Here, we will use $w = 200$ throughout, in line with \citet{xie_sequential_2013}.

\subsection{Monitoring by TPCA}
Algorithm \ref{alg:tpcaMonitoring} summarizes how TPCA is used in conjunction with the mixture monitoring procedure \eqref{eq:MixCorrectedStoppingRule} to solve the original
change-point detection problem \eqref{eq:OriginalHypothesis}.
Observe that the monitored observations are the standardized projections;
\begin{equation}
  z_{j, t} = \hatbv_j^\tp \bS_0^{-1}(\bx_t - \hatbmu_0) \big/ \sqrt{\hatlambda_j},
\end{equation}
for $j \in \mathcal{J}$, 
where $\hatbmu_0$ is the training sample mean, $\bS_0$ is the diagonal matrix of training sample standard deviations,
and $\{\hatlambda_j, \hatbv_j\}_{j \in \mathcal{J}}$ is the sample eigensystem of the training sample correlation matrix.
In other words, the observations $\bx_t$ are first standardized by $\bu_t = \bS_0^{-1}(\bx_t - \hatbmu_0)$,
since PCA is not invariant to scaling.
These standardized observations then form the basis of PCA, and we get the projections $y_{j, t} = \hatbv_j^\tp \bu_t$. 
Lastly, the projections are normalized by $z_{j, t} = y_{j, t} \big/ \sqrt{\hatlambda_j}$.
The reason for normalizing the projections is numerical stability, since the variance of $y_{j, t}$ for $j$ close to $D$ will typically be very small.

\begin{algorithm} 
  \caption{Monitoring by TPCA} \label{alg:tpcaMonitoring}
  \begin{algorithmic}[1]
    \INPUT{$b$, $p(\bmu_1, \bSigma_1 | \bSigma_0)$ and $\{ \bx_s \}_{s = -m + 1}^0$.} % assumed to come from $N(\bmu_0, \bSigma_0)$}
    \State Compute $\hatbmu_0$, $\bS_0$ and the correlation matrix $\hatbSigma_0$ from $\{ \bx_s \}_{s = -m + 1}^0$.
    %\State $\hatbmu_0 \gets \frac{1}{m} \sum_{s = -m + 1}^0 \bx_s$
    %\State $\hatbsigma_0 \gets \frac{1}{m - 1} \sum_{s = -m + 1}^0 (\bx_s - \hatbmu_0)^2$ and $\bS_0 \gets \diag\{ \hatbsigma_0 \}$.
    %\State $\bU_m \gets \bS_0^{-1}(\bX_m - \hatbmu_0)$.
    %\Comment Standardizing data before PCA.
    %\State $\hatbSigma_0 \gets \frac{1}{m - 1} \bU_m \bU_m^\tp$.
    \State $\mathcal{J}$ and $\{\hatlambda_j, \hatbv_j\}_{j \in \mathcal{J}} \gets $ the results of applying Algorithm \ref{alg:TailoredPCA} to $\hatbSigma_0$ with $p(\bmu_1, \bSigma_1 | \hatbSigma_0)$.
    \State $z_{j, t} \gets \hatbv_j^\tp \bS_0^{-1}(\bx_s - \hatbmu_0) \big/ \sqrt{\hatlambda_j}$, for $t = -m + 1, \ldots, 0$ and $j \in \mathcal{J}$.
    \State $t \gets 0$ and $\Lambda^C_{\text{max}, t}(1) = 0$.
    \While{$\Lambda^C_{\text{max}, t}(1) < b$}
    \State $t \gets t + 1$ and new data $\bx_t$ arrives. %, either from $N(\bmu_0, \bSigma_0)$ or $N(\bmu_1, \bSigma_1)$.
%     \State $\bu_t \gets \bS_0^{-1}(\bx_t - \hatbmu_0)$.
    \State $\bz_t \gets (z_{j, t}) \gets \hatbv_j^\tp \bS_0^{-1}(\bx_t - \hatbmu_0) \big/ \sqrt{\hatlambda_j}$ for $j \in \mathcal{J}$.
    \State $\Lambda^C_{\text{max}, t}(1) \gets \underset{k \in \mathcal{K}}{\max}\; \Lambda^C_{k, t}(1)$ based on $\{ \bz_s \}_{s = -m + 1}^t$.
    \EndWhile
    \myRETURN{$t$}
  \end{algorithmic}
\end{algorithm}

It is important to note that the estimates $\hatbmu_0$, $\bS_0$ and $\{\hatlambda_j, \hatbv_j\}_{j \in \mathcal{J}}$ are not updated as more data arrives.
Ideally they would be updated for every new observation $\bx_t$, but then the procedure would lose its sequential nature;
all projections $z_{j, s}$ for all $s$ would have to be recalculated at every step, as well as everything based on them.
Estimating the quantities needed for the projections only once in combination with incorporating the estimation uncertainty when 
calibrating the threshold $b$ is a solution that allows for both recursive computations on the projections and control of false alarms under a correctly specified model.

\subsection{Controlling False Alarms}
How can one set the threshold $b$? As in regular hypothesis testing there is a trade-off between false positives and false negatives.
There are several sequential analogs, but recall that we use the probability of false alarm \eqref{eq:PFA} and the expected detection delay \eqref{eq:EDD}, 
respectively, motivated by the discussion in \citet{lai_sequential_1995}.
A threshold $b$ can now be found by choosing a segment length $n$ and a probability of false alarm $\alpha$, then solving $\alpha = \p^\infty[T(p_0, b) \leq n]$ for $b$.
The EDD of the stopping rules can then be compared, where the goal is as low EDD as possible.

% Recall that $\p^\kappa$ and $\E^\kappa$ denote probability and expectation under the model where there is a true change-point at $\kappa$.
% In particular, $\p^\infty$ and $E^\infty$ means probability and expectation under $H_0$.
% If $T(b)$ is a stopping rule that stops when some statistic reaches the threshold $b$, the PFA is defined as 
% \[
%   \p^\infty[T(b) \leq n]
% \]
% for a chosen segment length $n$ , and the EDD as
% \[
%   \underset{\kappa}{\sup}\; \E^\kappa[T(b) - \kappa | T(b) > \kappa].
% \]
% Now a threshold $b$ can be found by choosing a segment length $n$ and a probability of false alarm $\alpha$, then solving $\alpha = \p^\infty[T(b) \leq n]$ for $b$.
% The EDD of the stopping rules can then be compared, where the goal is as low EDD as possible.
% 
\begin{myremark}
  The \textit{average run length} (ARL), defined as $\E^\infty[T(b)]$, is perhaps a more commonly used measure of false alarms.
  However, in many applications, a false alarm is very undesirable, and \citet{lai_sequential_1995} argues that a 
  more informative measure of false alarms is to consider the probability of no false alarm during a typical, steady-state period of operation.
  For example, an average run length of $1000$ does not necessarily mean that the probability of a false alarm during the first $100$ observations is low.
  Another advantage is that the PFA is much more tractable to compute by Monte Carlo simulation.
  Finally, also pointed out by \citet[p.~631]{lai_sequential_1995}, the two quantities are related approximately by
  \[
    \E^\infty[T(b)] \approx n / \p^\infty[T(b) \leq n],
  \]
  for the stopping rule we consider.
\end{myremark}

Finding thresholds for monitoring the raw data can be done by a relatively straight forward bootstrap procedure.
Thresholds for monitoring the PCA projections, however, are slightly more complicated to attain, so this is what we focus on below.
The accompanying R package \texttt{tpcaMonitoring} can be consulted for all implementational details.

Complications arise for monitoring the projections because uncertainty due to estimating the principal axes from the training data has to be incorporated.
If not, there will be false alarms due to estimation error rather than an actual change in the distribution.
Importantly, the estimation variance of the sample principal components can not necessarily be disregarded even for high sample sizes.
This is seen from the asymptotic distribution of the eigenvectors of a sample covariance matrix $\bSigma$.
Recall that $\{ \lambda_j, \bv_j \}_{j = 1}^d$ and $\{ \hatlambda_j, \hatbv_j \}_{j = 1}^d$ are the population and sample eigensystems, respectively.
Then $\hat{\mathbf{v}}_j$ is asymptotically multivariate normal with mean $\mathbf{v}_j$ and covariance matrix
\begin{equation}
  \bGamma_j = \frac{\lambda_j}{n} \sum_{l \not= j} \frac{\lambda_l}{(\lambda_j - \lambda_l)^2} \bv_j \bv_j^\tp,
\end{equation}
given that the $\lambda_j$'s are all distinct eigenvalues \citep[p.~405]{muirhead_aspects_1982}.
Hence, if there is a small gap between two population eigenvalues, the variance can be large even for large sample sizes.

The estimation uncertainty can be incorporated by the following bootstrapping procedure:
\begin{enumerate}
  \item Input: Training data $\{ \bx_s \}_{s = -m + 1}^0$ assumed to be $N(\bmu_0, \bSigma_0)$, $b$, $n$ and $\alpha$.
  \item Obtain estimates $\hatbmu_0$ and $\hatbSigma_0$ from the training data.
  \item Run the TPCA algorithm (Algorithm \ref{alg:TailoredPCA}) on $\hatbSigma_0$ to get the indices $\mathcal{J} \in \{ 1, \ldots, D \}$.
  \item Draw a bootstrap training sample $\{ \tildebx_s \}_{s = -m + 1}^0$, where $\tildebx_s \iidsim N(\hatbmu_0, \hatbSigma_0)$.
  \item Run Algorithm \ref{alg:tpcaMonitoring} on $\{ \tildebx_s \}_{s = -m + 1}^0$ and equally distributed monitoring observations $\tildebx_t \iidsim N(\hatbmu_0, \hatbSigma_0)$, $t = 1, \ldots, n$.
    %The TPCA algorithm is not rerun on the bootstrap training sample since it is the uncertainty in the eigenvectors $j \in \mathcal{J}$ that we want to capture.
    One exception is that $\mathcal{J}$ of $\hatbSigma_0$ is reused to select projections.
    %The output is a draw $T(1) = t$ if $T(1) \leq n$ or $T(1) = n + 1$ if $T(1) > n$.
  \item Record $I\{T(1, b) \leq n\}$.
  \item Repeat 4 - 6 B times.
  \item Average the B $I\{T(1, b) \leq n\}$'s to get an estimate $\hatalpha$ of $\p^\infty[T(1, b) \leq n]$.
\end{enumerate}
Finally, repeat for different $b$'s until $\hatalpha$ is close enough to $\alpha$ within a desired margin of error.
%This bootstrap procedure mimics TPCA monitoring under $H_0$, but takes into consideration that the training set $\bX_m$ is a random matrix in addition to the monitoring observations.

This parametric bootstrap procedure also opens the door for other ways of robustifying the threshold;
pick another distribution to bootstrap training and monitoring samples from than the normal, and run the otherwise exact same simulations.
For example the multivariate t-distribution or the empirical distribution function (i.e., a nonparametric bootstrap).

A drawback of using bootstrapping to get a threshold $b$ for TPCA monitoring is that each threshold is conditional on the exact training set, 
which principal axes $\mathcal{J}$ as well as the window size $w$.
Thus, to obtain exact error control under the aassumption of a correctly specified model, a new threshold must be found by simulation for every training set.
Luckily, these simulations depend most strongly on the number of projections $|\mathcal{J}|$ rather than $D$, making it scalable.
Setting up and running these simulations is the cost of incorporating all sources of uncertainty and achieving exact error control.
% 
% Drawbacks:
% \begin{itemize}
%   \item A threshold $b$ for tpca monitoring is conditional on the exact training set, chosen principal axes $\mathcal{J}$ as well as the window size $w$.
%   \item The raw mixture procedure depends on the dimension of the training set $m$ and $d$, as well as $p_0$ and $w$.
%   \item For high $D$, computationally heavy. Does not scale well. FALSE. Depends on $r$ in stead of $d$.
% \end{itemize}
% 
% \printbibliography

\section{Numerical Performance Analysis} \label{sec:5}
In this section, the detection performance of TPCA monitoring is assessed through an extensive simulation study.
The three questions we want to answer are: In terms of EDD, how well does TPCA monitoring compare to another method that also explicitly handles sparse changes?
What is gained by using TPCA compared to simply picking the least varying projections?
How much can the dimension be reduced by without compromising on detection speed?

\subsection{Setup}
In all the simulations we present here, $n = 100$ and $\alpha = 0.01$ with $95\%$ confidence (an ARL of approximately $10^4$) and $w = 200$.
The main simulation study is performed for $D = 100$ and $m = 200$, while some results for $D = 500$ with $m = 1000$ are presented briefly at the end of the results section.
Four different classes of methods were run on each change scenario: The mixture procedure on the raw data with method parameters $p_0 = 0.03, 0.1, 0.3, 1$,
the $J = 1, 2, 3, 5, 10, 20$ most (Max PCA) and least (Min PCA) varying projections, as well as TPCA with cutoffs $c = 0.8, 0.9, 0.95, 0.99, 0.995, 0.999$.
For TPCA, change distribution \eqref{eq:ex_change_distribution} was used with some modifications to see if incorporating information had any effect.
To be precise, we assumed knowledge about which change type was of interest, so for changes in mean, for example, we set $p_\mu = 1$ and the others to $0$.
In addition, we set $K \leq D/2$ to emphasize sparse changes.
% Tables \ref{tab:dense_J}, \ref{tab:halfsparse_J} and \ref{tab:sparse_J} in the Appendix can be consulted to see which axes were chosen.
% Table \ref{tab:dense_J} in the Appendix can be consulted to see which axes were chosen.

\begin{figure}[htb]
\centering
\includegraphics[scale = 0.7]{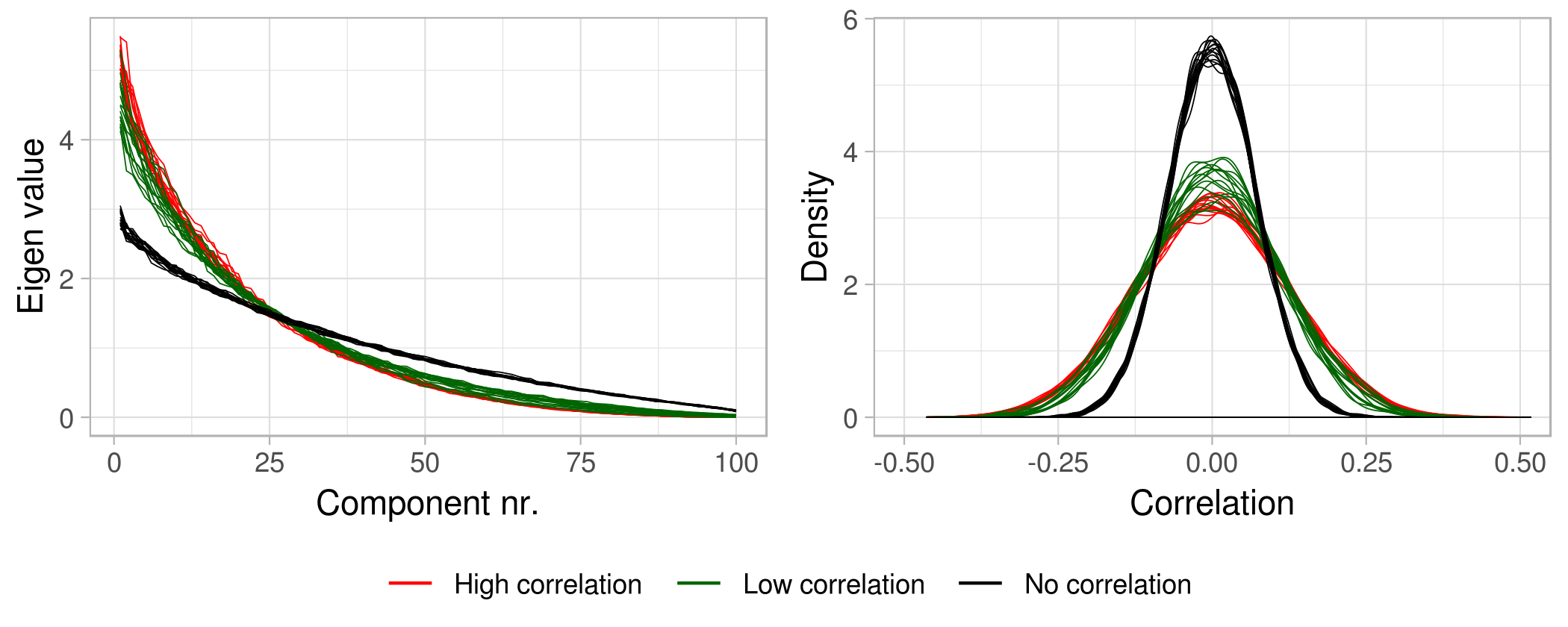}
\caption{Scree plots (left) and corresponding correlation density plots (right) of the $30$ random training $\hatbSigma_0$'s used in the simulations. 
As a reference for the spread of the matrices, 15 estimates based on 200 standard normal samples are also shown in black.}
\label{fig:scree}
\end{figure}

All the different change scenarios were considered for $30$ randomly chosen pre-change correlation matrices $\bSigma_0$, with varying strengths of correlation.
For each correlation matrix, a training set of $m = 200$ observations was drawn independently from $N(\mathbf{0}, \bSigma_0)$.
$15$ matrices fall into a "low correlation" group and $15$ into a "high correlation" group (Figure \ref{fig:scree}).
"Low" and "high" refers to different choices of the $\alpha_d$ parameter in the method of \citet{joe_generating_2006} for generating random correlation matrices,
where $\alpha_d < 1$ ($\alpha_d > 1$) yields a higher (lower) probability of large correlations in the space of correlation matrices.
The $\alpha_d$'s are evenly spread between 1 and 50 in the "low" group, while between 0.05 and 0.95 in the "high" group.
% All the different change scenarios were considered for a randomly chosen pre-change correlation matrix $\bSigma_0$ with sparsities $K_0 = 10$ (sparse), $50$ (half-sparse) and 100 (dense).
% To illustrate the spread of the different training sets, Figure \ref{fig:scree} shows the scree plots for the empirical covariance matrices together with a 
% reference plot based on a normal sample drawn with an identity covariance matrix.
%Because thresholds for TPCA are conditional on the exact training set, this training set was kept constant through all simulations.

After a change-point at $\kappa = 0$, observations to monitor were drawn independently from $N(\bmu_1, \bSigma_1)$.
For all change types and sizes, the proportion of affected streams was varied over $p = |\mathcal{D}|/D = 0.02, 0.05, 0.1, 0.3, 0.5, 0.7, 0.9, 0.95, 0.98$.
Which $|\mathcal{D}|$ dimensions that were changed was (uniformly) randomized in every simulated change.
Considered changes in the mean were $\mu_d = 0.5, 0.7, 1, 1.3$ for $d \in \mathcal{D}$, where $\bSigma_1 = \bSigma_0$.
To explain the changes in variance, note that any covariance matrix $\bSigma$ can be decomposed into its variance and correlation part by $\bSigma = \mathbf{C} \mathbf{R} \mathbf{C}$,
where $\mathbf{R}$ is the correlation matrix corresponding to $\bSigma$, and $\mathbf{C}$ is a diagonal matrix with the standard deviations $\bsigma$ on its diagonal.
Using this relationship, keeping the mean and the correlation matrix constant, the affected standard deviation components were changed to $\sigma_d = 0.5, 0.75, 1.5, 2$.
Finally, correlations $\rho_{di}$ were changed multiplicatively by factors $a_{di} = 0, 0.25, 0.5, 0.75$ for $d \not= i \in \mathcal{D}$,
while $\bmu_1 = \bmu_0$ and $\bsigma_1 = \bsigma_0$.
% Some of these changes in correlation result in indefinite matrices, and we explain a solution in the supplementary material.

In total, the setup consists of a grid of 108 change scenarios (combinations of change type, change size and change sparsity).
500 simulations of each change scenario is performed, and all the 22 combinations of methods and method parameters ($p_0$, $J$ or $c$) are run on every simulated data set to estimate the EDD.
Finally, everything is repeated for each of the 30 training sets, including finding new thresholds for all the PCA-based method.
This is important to have in mind to grasp the upcomming figures and results, which are compact summaries of a vast amount of simulations.
Also note that the figures showing EDDs have $log(p)$ on the x-axis to highlight the sparse change scenarios.

% Experimental setup for the EDD comparisons.
% * alpha and n: 0.01, n = 1000 (-> ARL: approx 10^5).
% * w: 200
% * D and m, types of bSigma_0: 100, 300, dense (K_0 = 100), halfsparse (K_0 = 50), sparse (K_0 = 10)
% * Input parameters for plain mixture: $p_0 = 0.03, 0.1, 0.3, 1$.
% * Input parameters for PCA methods: $J = 2, 3, 5, 10, 20$ and $c = 0.8, 0.9, 0.95, 0.99, 0.995, 0.999$
% * Which axes were selected for TPCA?

\subsection{Results}
\begin{figure}[htb]
\centering
\includegraphics[scale = 0.55]{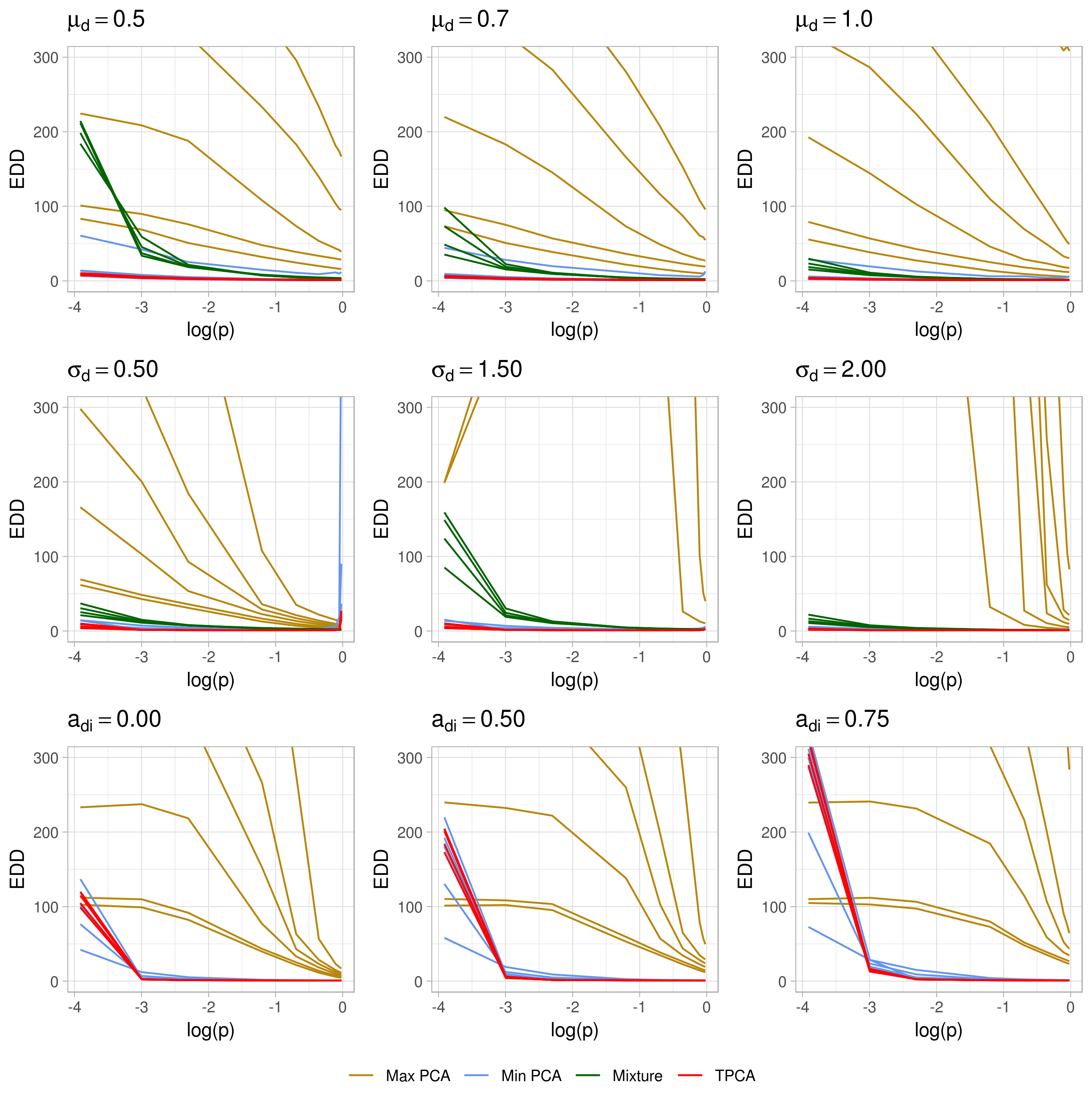}
\caption{\textbf{High correlation:} EDD for changes in the mean (first row), variance (second row) and correlation (third row) of varying change sparsity and change size.
Each line shows the EDD based on 500 simulations for a single method parameter, averaged over the 15 full runs with different high-correlation training sets.
Note that the mixture procedure applied to the raw data can not detect changes in correlation and is thus absent from the last line of figures.}
\label{fig:EDDhigh}
\end{figure}

\begin{figure}[htb]
\centering
\includegraphics[scale = 0.55]{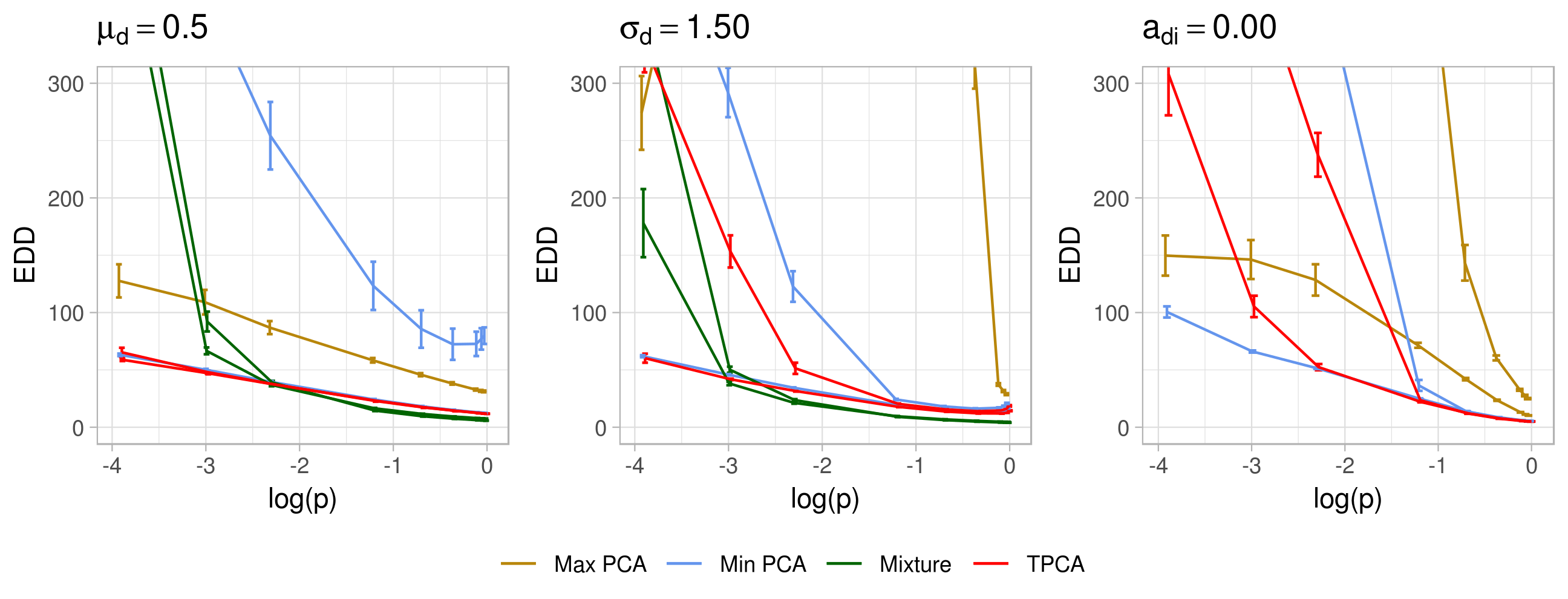}
\caption{An illustration of the errors involved in Figures \ref{fig:EDDhigh} and \ref{fig:EDDlow}:
Average EDDs with average 95\% confidence intervals for a subset of method parameters.
The averages are taken over all the 30 training sets, meaning that each confidence limit is an average of 30 individual limits.
EDD estimates that are high or of sparse changes are generally more uncertain.}
\label{fig:EDD_ci}
\end{figure}

When the correlations are high, monitoring the least varying projections through TPCA or Min PCA
can detect almost all the tested changes immediately with an EDD of 2-3 (Figure \ref{fig:EDDhigh} and \ref{fig:EDD_ci}, and Table \ref{tab:EDD_summary_high}).
Particularly, even the sparsest ($p = 0.02$), smallest changes in the mean and variance ($\mu_d = 0.5$ and $\sigma_d = 0.75, 1.5$) can be detected at this speed by monitoring only the two least varying projections.
I.e., a dimension reduction of $98\%$ can be obtained, while gaining in detection speed compared to the mixture procedure.
The sparsest changes in correlation is the only notable exception, where the EDD is 100-300 time-steps, depending on the size of the change.
Monitoring the most varying projections leads to considerably slower detection.

% Mon Apr  8 20:03:45 2019 
\begin{table}[htb] 
\caption{\textbf{High correlation:} Average EDD per change type for each method's best method parameters (in parenthesis), as a summary of Figure \ref{fig:EDDhigh}.
The average is taken over change sparsity, change size and the 15 full runs with different training sets.
To display robustness, the listed method parameters are the ones that are within 1 time unit of the method's minimum average EDD.} 
\label{tab:EDD_summary_high} 
\centering 
\begin{tabular}{lllll} 
\toprule 
 & \multicolumn{4}{c}{EDD} \\
\cmidrule{2-5} 
Change type & Max PCA($J$) & Min PCA($J$) & TPCA$(c)$ & Mixture($p_0$)\\ 
\midrule 
Mean & 27.4 (20) & 1.6 (2, 3, 5, 10) & \begin{tabular}{@{}c@{}l@{}} \multirow{2}{*}{1.8} & \hspace{0.15cm}(0.8, 0.9, 0.95, \\ & \hspace{0.15cm} 0.99 0.995, 0.999) \end{tabular} & 15.2 (0.03, 0.1) \\ 
Variance & 198.9 (20) & 2 (2, 3, 5) & \begin{tabular}{@{}c@{}l@{}} \multirow{2}{*}{2.1} & \hspace{0.15cm}(0.8, 0.9, 0.95, \\ & \hspace{0.15cm} 0.99 0.995, 0.999) \end{tabular} & 8 (0.03) \\ 
Correlation & 50.2 (20) & 10.8 (20) & 22.4 (0.999) & \\ 
\bottomrule 
\end{tabular} 
\end{table}

\begin{figure}[htb]
\centering
\includegraphics[scale = 0.55]{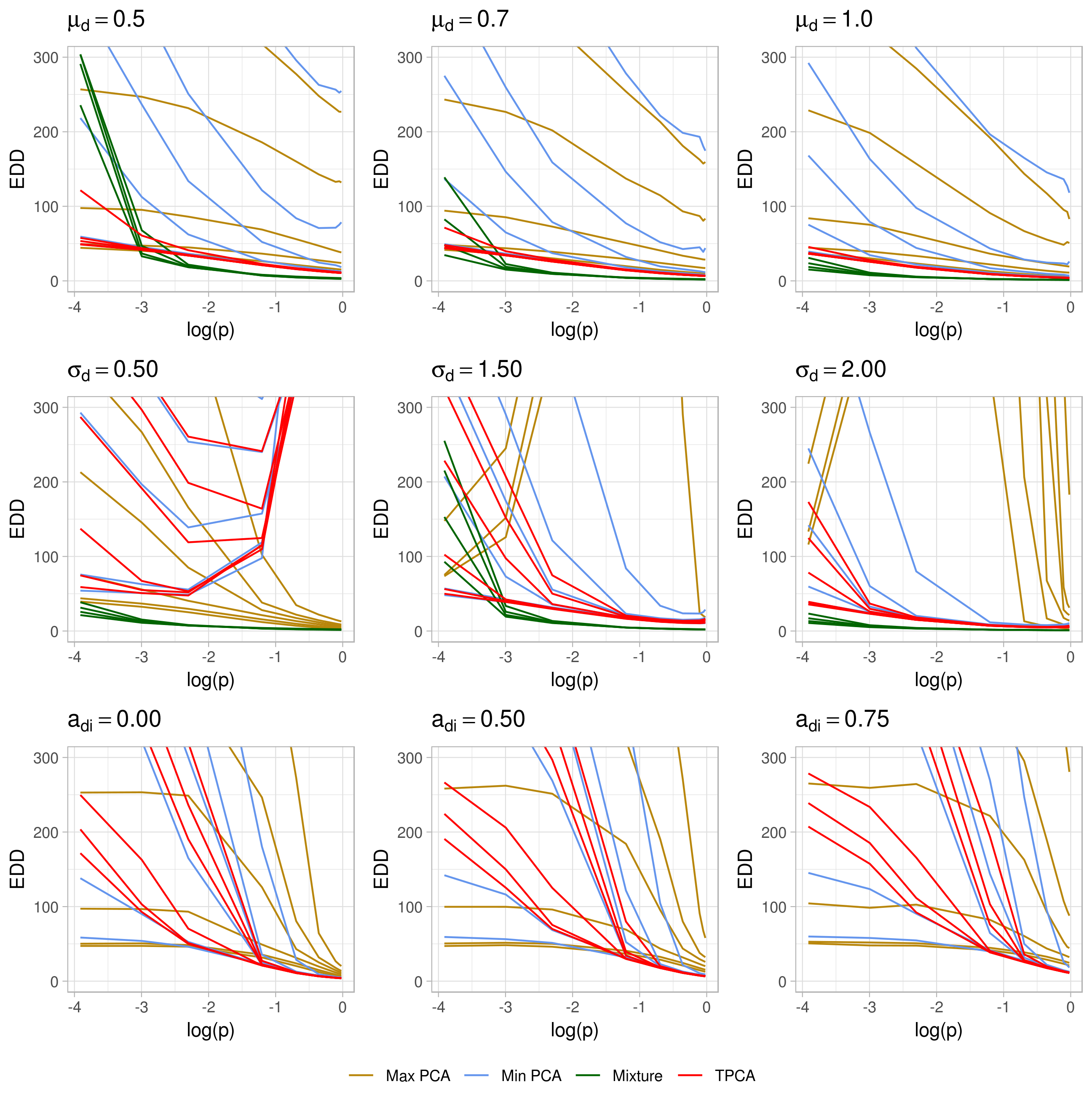}
\caption{\textbf{Low correlation:} EDD for changes in the mean (first row), variance (second row) and correlation (third row) of varying change sparsity and change size.
Each line shows the EDD based on 500 simulations for a single method parameter, averaged over the 15 full runs with different low-correlation training sets.
Note that the mixture procedure applied to the raw data can not detect changes in correlation and is thus absent from the last line of figures.}
\label{fig:EDDlow}
\end{figure}

As the correlations between streams become smaller, the performance of the least varying projections deteriorate (Figure \ref{fig:EDDlow} and \ref{fig:EDD_ci}, and Table \ref{tab:EDD_summary_low}).
In general, 10-20 projections are needed to attain a comparable performance with the mixture procedure; slightly worse performance for the denser, larger changes and better for the the sparse, small changes.
Thus, when the correlations are low, some compromise on detection speed must mostly be made to reduce the dimension, 
but a reduction of $80-90\%$ will often bring the EDD within 10 time units of the mixture procedure.
The most noticable difference from the high correlation scenario occurs when the variance decreases, where the most varying projections now are the most sensitive.
Note that this behaviour is in line with the two-dimensional results of \citet[p.~5]{tveten_which_2019}.
For changes in correlation when the correlations are small, we see that an even higher $c$ than $0.999$ is needed for TPCA to pick enough axes to detect the sparse changes as efficiently as Min and Max PCA with 20 projections.

% Mon Apr  8 20:04:52 2019 
\begin{table}[htb] 
\caption{\textbf{Low correlation:} Average EDD per change type for each method's best method parameters (in parenthesis), as a summary of Figure \ref{fig:EDDlow}.
The average is taken over change sparsity, change size and the 15 full runs with different training sets.
To display robustness, the listed method parameters are the ones that are within 1 time unit of the method's minimum average EDD.} 
\label{tab:EDD_summary_low} 
\centering 
\begin{tabular}{lllll} 
\toprule 
& \multicolumn{4}{c}{EDD} \\
\cmidrule{2-5} 
Change type & Max PCA($J$) & Min PCA($J$) & TPCA$(c)$ & Mixture($p_0$)\\ 
\midrule 
Mean & 20.6 (20) & 17.9 (20) & \begin{tabular}{@{}c@{}l@{}} \multirow{2}{*}{17.6} & \hspace{0.15cm}(0.9, 0.95, 0.99, \\ & \hspace{0.15cm} 0.995, 0.999) \end{tabular} & 16 (0.03) \\ 
Variance & 184.3 (20) & 158.8 (10) & 154.8 (0.995) & 8.3 (0.03) \\ 
Correlation & 30 (20) & 28.5 (20) & 52.3 (0.999) & \\ 
\bottomrule 
\end{tabular} 
\end{table}

In terms of detection speed alone, there is no advantage of using TPCA compared to simply picking the axes of the least varying projections as in Min PCA;
more or less the same projections are monitored under both schemes.
However, Tables \ref{tab:EDD_summary_high} and \ref{tab:EDD_summary_low} point to the fact that TPCA will automatically choose a reasonable subset of projections
quite robustly with respect to the cutoff $c$ given some knowledge about which changes are of interest.
This robustness is not observed to the same degree by picking projections manually with Min PCA.
The exception to this rule is for decreases in variance and changes in correlation of weakly correlated data.

\begin{figure}[htb]
\centering
\includegraphics[scale = 0.55]{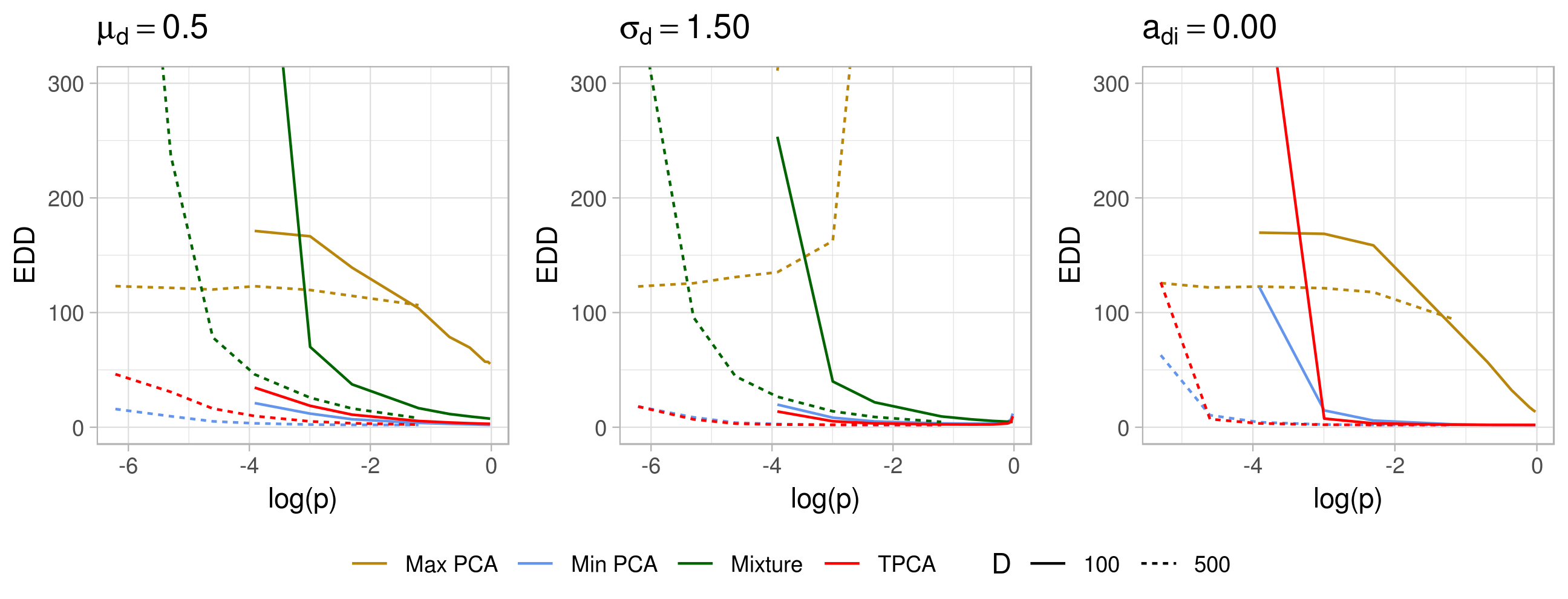}
\caption{A comparison of EDDs for $D = 100, 500$ with $m = 2D$ for a single training set generated from a randomly drawn correlation matrix using $\alpha_d = 1$.
Only the best choices of method parameters are shown. 
These figures are representative of the pattern also seen for $\mu_d = 0.7, 1$, $\sigma_d = 0.5, 2$ and $a_{di} = 0.5, 0.75$, which are omitted due to space limitations.}
\label{fig:EDD_compare}
\end{figure}

Lastly, a hint towards the generalizability of these results to higher dimensions than $D = 100$ is given in Figure \ref{fig:EDD_compare}.
For $D = 500$, the change sparsities tested was $p = 0.002, 0.005, 0.01, 0.02, \allowbreak 0.05, \allowbreak 0.1, 0.3$.
Observe that TPCA and Min PCA are still able to detect the very sparse changes in the 500-dimensional data stream at almost the same speed as the sparse ones in the 100-dimensional stream.

\section{Testing on the Tennessee Eastman Process} \label{sec:6}
% Highlight: Only training set. Nothing else. No extra validation set.
% 
In this section, TPCA is extended to handle time-dependent data and compared to dynamic PCA (DPCA) as used in the stochastic process control (SPC) litterature 
and the method of \citet{kuncheva_pca_2014} (what we have called Min PCA or Min DPCA below).
See for example \citet{vanhatalo_structure_2017} or \citet{rato_systematic_2016} for an introduction to DPCA in SPC.
The methods are tested on the well-known Tennessee Eastman Process (TEP): a model of an industrial chemical process used to generate realistic data from a large system \citep{downs_plant-wide_1993}.
As a test bed, we will use the available TEP dataset from \citet{rieth_additional_2017}, 
which includes fault free training sets of 500 observations as well as faulty test sets of 960 observations with a change-point at $\kappa = 160$.
Each observation is a sample from the process with 3 min intervals and consists of 41 direct measurements (xmeas) and 11 controlled input variables (xmv), 52 in total.
There are 500 complete test sets for each of 20 different faults.
Most faults cause sparse distributional changes, where faults 1-7 are changes in mean, fault 8-12 are changes in the variance, and the rest are of various other types \citep{rato_systematic_2016}.
As before, we measure the EDD of detecting these faults after a short training period, now on $m = 500$ observations.
We stress that this means no extra validation set is available for fine-tuning, which DPCA generally depends on.

To let TPCA account for the highly auto-correlated TEP observations, we extended it in similar fashion as PCA is extended to DPCA.
I.e., a lag $l$ is chosen, and the observation vectors $\bx_t$ are lag-extended to $\tildebx_t = (\bx_{t - l}^\tp, \ldots,  \bx_t^\tp)^\tp$ before they are fed to PCA.
This induces a VAR model with lag $l$ on the data.
Then, a change distribution for $\bx_t$ can be set up like before,
where each simulated change now corresponds to $l + 1$ duplicate changes in the parameters of $\tildebx_t$.
% but when simulating changes, the corresponding parameters of $\tildebx_t$ are chnged the block structure that is imposed on the mean vector and correlation matrix of $\tildebx_t$ must be taken into account.
% For example, simulating a change in the variance of one component of $\bx_t$ means that $l + 1$ variances of the lag-extended data $\tildebx_t$ are changed, and similarly for the other change types.
In this way, pre- and post-change parameters of $\tildebx_t$ are obtained, which can be used to measure the sensitivity of the projections $\tildebv_j^\tp \tildebx_t$,
where $\tildebv_j$, $j = 1, \ldots, D(l + 1)$, are the eigenvectors of the correlation matrix of $\{\tildebx_t\}_{t = - m + l + 1}^0$.
We call this method tailored dynamic PCA (TDPCA), and we will also compare it to simply picking the $J$ most (Max DPCA) and least (Min DPCA) varying projections, like previously.
It is also possible to implement a change distribution over changes in the auto-correlations, but for simplicity we keep using change-distribution \eqref{eq:ex_change_distribution} and close relatives.
When $p_\mu = 1$, $p_\sigma = 1$ and $p_\mu = p_\sigma = p_\rho = 1/3$, we denote the methods by TDPCA(mean), TDPCA(var) and TDPCA(unif), respectively.

The main additional challenge comes from setting a valid detection threshold when the observations are not independent multivariate normal.
We tackle this by switching the parametric bootstrap procedure of Section \ref{sec:4} with a non-parametric block bootstrap \citep{kunsch_jackknife_1989}.
Thresholds are set to meet a PFA of $\alpha = 0.01$ on $n = \kappa - l$ observations with $90\%$ confidence, for comparisons with the empirical probability of false alarms $\hat{\alpha}$.

\citet[p.~10]{vanhatalo_structure_2017} suggest lags $l = 2, 3$ or $5$ for DPCA on the TE process.
In our case, using lags $2$ and $3$ were not sufficient to capture most autocorrelation, so we proceeded with $l = 5$ for all methods as well as in the bootstrap.
This yields 312-dimensional lag-extended observations with a change-point at $\kappa - l = 155$.
% 
% The comparison between TDPCA and DPCA as used in stochastic process control is not entirely unproblematic.
% This is mainly because monitoring by TDPCA is based on change-point detection methodology, trying to identify changes in the underlying parameters of the data.
% On the other hand, DPCA uses the Hotelling's $T^2$ and $Q$-statistic, which only measures how non-conforming each new data point is to the trained DPCA model. 
% In practice, however, these methodologies will often be used for the same purpose (detecting faults of some kind), which is why we find the comparison fruitful.

Before proceding to the results, DPCA must be fit into the framework of controlling the PFA and measuring EDD.
DPCA is not a change-point method, and only measures how non-conforming each new data point $\tildebx_t$ is to the trained DPCA model by two statistics.
The Hotelling's $T^2(\tildebx_t)$ is the squared Mahalanobis distance of $\tildebx_t$ in the DPCA model subspace, 
and the $Q(\tildebx_t)$-statistic measures the orthogonal distance of $\tildebx_t$ to the DPCA subspace \citep{rato_systematic_2016}.
The corresponding stopping rule for DPCA is $T_\text{DPCA} = \min(T_1, T_2)$, where 
\[
  T_1 = \inf\{ t \geq 0 : T^2(\tildebx_t) > T^2_{\alpha_n} \} \text{ and } T_2 = \inf\{ t \geq 0 : Q(\tildebx_t) > Q_{\alpha_n} \} \},
\]
$T^2_{\alpha_n}$ and $Q_{\alpha_n}$ being percentiles of each statistic's distribution.
The percentiles depend on $n$ to fulfill the PFA requirement $\p^\infty(T_\text{DPCA} \leq n) \leq \alpha$.
By assuming that false alarms are equally likely for both statistics and applying a union bound, we get that $\alpha_n = \alpha / (2n)$.
This is a slightly conservative percentile.
Since we do not have a validation set to find the two thresholds precisely, we use the common approximation of $T^2$ being $\chi^2_r$, 
where $r$ is the number of retained components in the DPCA model, and the approximation of the $Q$-statistic given by \citet{jackson_control_1979}.
Since the approximation for the $T^2$ assumes normality and temporal independence of the retained (most varying) projections, we expect it to be overly optimistic for the TEP data,
but counter-weighted somewhat by the conservative bound on $\alpha_n$.
Thus, it is a realistic setup in the case of no validation set, but we don't expect it to work very well in terms of false alarm control.

Lastly, in the results below, we have set the cumulative percentage of variance explained in DPCA to $95\%$,
$J = 20$ in Min and Max DPCA, $c = 0.9$ in TPCA(mean) and TPCA(uniform), and $c = 0.99$ in TPCA(var).
The reason for the cutoff-values is that these also result in approximately 20 projections being chosen, to be comparable with Min and Max PCA.
Note that it is generally better to set $c$ too high than too low, as too few projections being chosen can be detrimental, 
while including a few more projections than necessary only slows detection slightly.
Our experiments suggest that a dimension reduction of $7-10\%$ is a good choice.

% By construction, we expect that smaller changes will be detected quicker by TDPCA, while larger changes will be detected quicker by DPCA, 
% and that TDPCA will be more robust to false alarms.

\begin{figure}[htb]
\centering
\includegraphics[scale = 0.5]{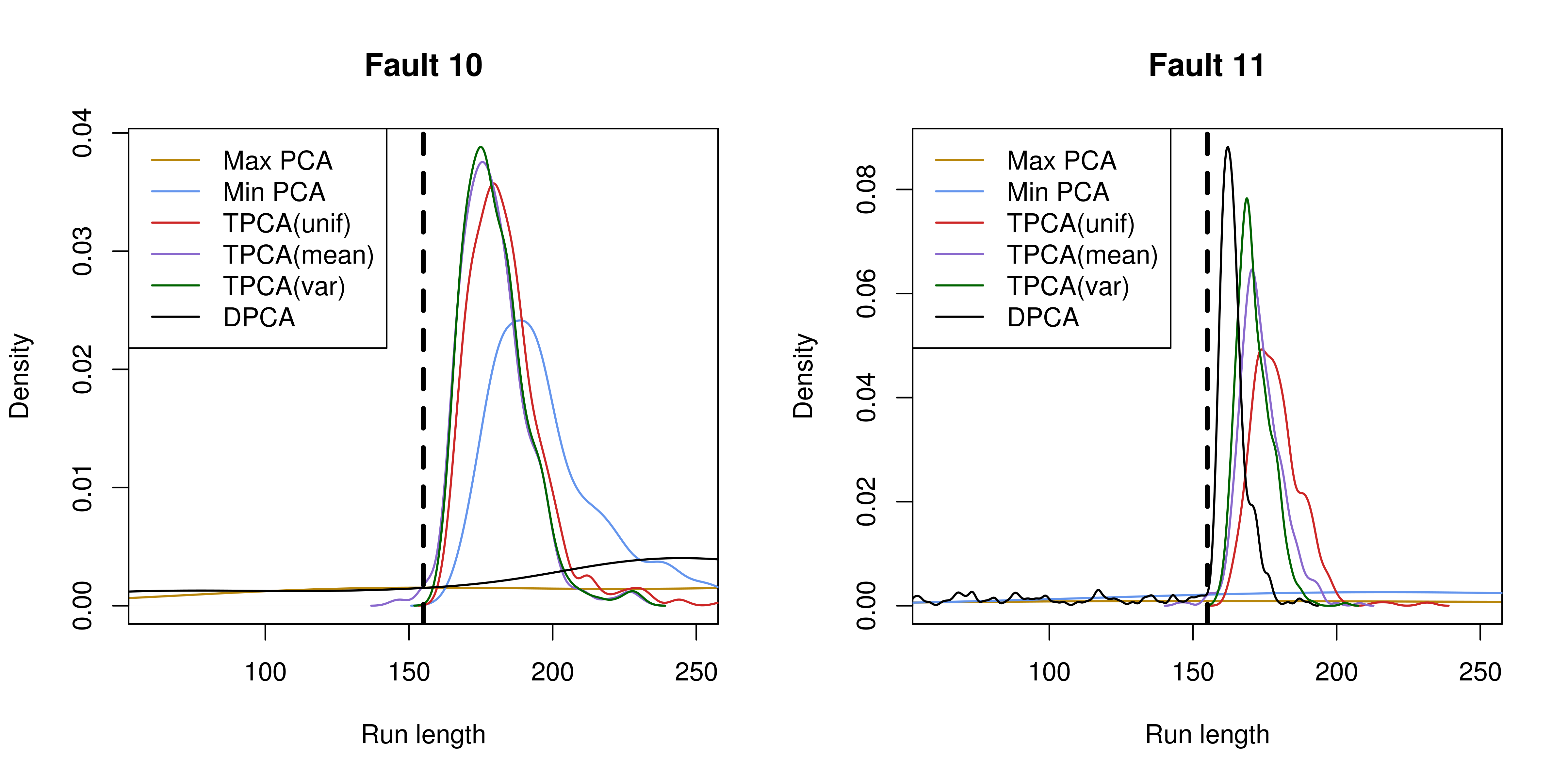}
\caption{Kernel density estimates of run lengths for faults 10 and 11 of the TE process. The dashed line marks the change-point.}
\label{fig:TEP_densities}
\end{figure}

The results are summarized in Table \ref{tab:TEP} and Figure \ref{fig:TEP_densities}.
As expected, the proportions of false alarms for DPCA is much higher than the nominal $0.01$, which disqualifies it for use in this setting.
The same is the case for Max DPCA, which suffers from the most varying projections being long-range auto-correlated.
Among the methods that achieve appropriate error control, one of the TDPCA variants are the quickest in all cases.
In particular, note that TDPCA still beats DPCA in 13 out of 20 cases despite the considerably stricter control on false alarms.
When we gave DPCA the luxury of a massive validation set to find more accurate thresholds, we still found that TDPCA beats DPCA in 15 out of 20 cases.
Unexpectedly, there is no systematic relation between the type of change and whether TDPCA(mean) and TDPCA(var) is best, they are mostly almost equal, and only slightly faster than TDPCA(unif).
Given the results of Section \ref{sec:5}, it is slightly surprising that TDPCA is significantly better than Min DPCA.

% Table \ref{tab:TEP} and Figure \ref{fig:TEP_densities} show that the probability of false alarms is controlled quite well by the block bootstrap procedure if the least varying projections are selected,
% which is mostly the case for Min PCA and TPCA using change distribution \eqref{eq:ex_change_distribution} with $p_\sigma = 1$ (TPCA (var) in the table).
% Simultaneously, rather quick detection of the sparse changes of both fault 4 and 11 can be achieved (Min PCA(J = 20) and TPCA (c = 0.99)).
% If some of the more varying projections are included, on the other hand, the error control is inaccurate.
% A likely reason is that the more varying projections are much more strongly auto-correlated than the less varying ones in the training set.
% This is a consequence of high auto-correlation not captured by the VAR(3) model being translated into high variance when PCA is performed on the lag-extended data.
% Both TPCA with all change types being equally likely (TPCA (uniform)) and $p_\mu = 1$ (TPCA (mean)) selects some projections in the middle that are still quite strongly auto-correlated.
% An explanation for this choice of axes is that TPCA optimizes the choice of projections to be the most sensitive to distributional changes, an opposing force to being robust to model misspecification.

% latex table generated in R 3.5.0 by xtable 1.8-2 package
% Tue May  7 11:26:20 2019
\begin{table}[ht]
\caption{EDD results for the TEP data. The quickest methods among those that achieve acceptable error control ($\hat{\alpha} \leq 0.01$) are in bold.}
\label{tab:TEP}
\centering
\begin{tabular}{rrrrrrr}
  \toprule
  Fault & Min DPCA & TDPCA(unif) & TDPCA(var) & TDPCA(mean) & Max DPCA & DPCA \\ 
  \midrule
  $\hat{\alpha}$ & 0.000 & 0.000 & 0.000 & 0.008 & 0.156 & 0.182 \\
  \midrule
  1 & 14.6 & 7.4 & 7.4 & \bf{5.4} & 17.0 & 6.2 \\ 
  2 & 42.4 & 22.8 & 19.7 & \bf{17.0} & 25.3 & 13.9 \\ 
  3 & 800.0 & 769.4 & 695.5 & \bf{663.4} & 668.5 & 416.7 \\ 
  4 & 757.8 & 20.3 & \bf{9.6} & 12.9 & 127.9 & 1.0 \\ 
  5 & 2.9 & 2.0 & \bf{1.8} & 2.6 & 27.0 & 2.1 \\ 
  6 & 1.9 & 1.2 & \bf{1.0} & \bf{1.0} & 23.3 & 1.0 \\ 
  7 & 5.1 & 2.6 & 2.7 & \bf{1.9} & 9.8 & 1.0 \\ 
  8 & 49.0 & 24.7 & 24.2 & \bf{20.0} & 40.7 & 23.0 \\ 
  9 & 800.0 & 766.4 & 695.6 & \bf{661.8} & 663.4 & 401.6 \\ 
  10 & 45.0 & 28.6 & 25.0 & \bf{24.6} & 317.0 & 158.0 \\ 
  11 & 308.0 & 24.0 & \bf{16.3} & 18.6 & 635.9 & 9.2 \\ 
  12 & 9.1 & 7.8 & \bf{7.3} & 7.8 & 34.6 & 9.1 \\ 
  13 & 81.7 & 47.2 & 43.3 & \bf{39.6} & 61.6 & 45.5 \\ 
  14 & 54.2 & 27.0 & 22.4 & \bf{20.1} & 9.7 & 2.3 \\ 
  15 & 800.0 & 313.7 & \bf{80.8} & 208.3 & 672.3 & 377.1 \\ 
  16 & 31.5 & 17.7 & 16.3 & \bf{15.5} & 603.4 & 219.3 \\ 
  17 & 44.9 & 36.4 & 34.5 & \bf{33.5} & 46.5 & 35.0 \\ 
  18 & 60.4 & 51.6 & 48.7 & \bf{48.0} & 73.8 & 51.6 \\ 
  19 & 723.2 & 14.6 & 13.5 & \bf{9.8} & 692.1 & 19.1 \\ 
  20 & 52.9 & 43.5 & \bf{38.4} & 40.0 & 505.0 & 48.2 \\ 
   \bottomrule
\end{tabular}
\end{table}

% \section{Discussion} \label{sec:7}
% Validity of results: Only three covariance matrices and training sets, can general conclusions be made?
% Robustness to the normality assumption.
% What are the advantages of TPCA over Min PCA?
% How will another change distribution affect the results?
% Post-change diagnosis.

\section{Conluding Remarks} \label{sec:8}
The problem of detecting sparse changes in the mean and/or covariance matrix of high-dimensional data is a problem that admits no efficient direct solution because of the number of samples
necessary to estimate the covariance matrix.
Monitoring projections of the incoming data onto the pre-change principal axes offers an indirect solution that is also computationally scalable.
Which projections to monitor for specific distributional changes is not self-evident, and this choice is what TPCA offers an answer to.
We have seen that TPCA's choice of projections work well in almost all cases studied when modelling assumptions are correct,
the exception being sparse changes in the correlation, and decreases in variance when the correlations are weak.
Monitoring the TPCA projections work especially well if the data streams are strongly correlated, where most changes, even very sparse and small ones, 
can be detected almost immediately in 100-dimensional normal data.
On the other hand, if correlations are weak, some performance is lost by dimension reduction, 
but one still gains the ability to detect changes in correlation without loosing too much in detection speed for changes in the mean and variance compared to the benchmark mixture procedure.

On the TEP data, we saw that the dynamic version of TPCA combined with a non-parametric block-bootstrap procedure to robustly set thresholds worked well in detecting a wide range of changes quickly.
Importantly, this was achieved without a large validation set, which is often needed to make the classic SPC tool DPCA work properly.
For error control to be achieved, however, enough lags must be included so that the TDPCA projections are not subject to major auto-correlation.
In terms of detection speed, TDPCA improves upon the method of \citet{kuncheva_pca_2014} (Min DPCA), and is also slightly quicker than DPCA in most cases.
The superiority of TDPCA over DPCA is most notable when the changes are small and sparse, whereas most changes in the TEP data are sparse, but large.

% When modelling assumptions are violated, TPCA is not guaranteed to select a suitable subset, so more care must be taken during the training phase.
% Nevertheless, the bootstrapping approach for setting detection thresholds presented here robustifies the procedure, shown by the example of the Tennessee Eastman process;
% unless very strong model deviations are present in the chosen projections, the bootstrap approach can achieve false alarm control that is not overly pessimistic.
% %For example, a strong model deviation is more than the data actually coming from an $AR(2)$ model with a moderate auto-correlation when it is modelled as i.i.d. normal.
% An option for the future is to develop another criteria to select projections by that maximizes a more robust quality than sensitivity to changes,
% since sensitivity to changes of parameters of a given model also will mean sensitivity to all forms of model deviations.

It should also be mentioned that we have only considered the question of when to raise an alarm.
After an alarm has been raised, it is of course natural to ask which parameters and which dimensions/sensors that changed.
This question is left for future research, but relevant litterature already exists in e.g. \citet{hawkins_multivariate_2009} and \citet{lakhina_diagnosing_2004}.

Other interesting follow-up questions include: (1) How does TPCA work combined with more sophisticated tools for handling time-dependence?
(2) We have studied a general pre-change covariance matrix. What if it has a known structure of a certain form, like blockwise-dependence?
(3) Can the insight about PCA for change detection provided here be extended to other dimension reduction tools?

\section*{Acknowledgements}
This work is partially funded by the Norwegian Research Council centre Big Insight, project 237718.
We would also like to thank Kristoffer H. Hellton and Steve Marron for useful discussions regarding the theory of PCA.

\section*{Supplementary Materials}
\begin{description}
  \item[Appendix:] A) Sensitivity under other change distributions, B) Dealing with indefinite post-change correlation matrices. (.pdf document)
  \item[R-package \texttt{tpca}:] The TPCA routine for selecting projections (Algorithm \ref{alg:TailoredPCA}). 
                                  Also includes the dynamic version of TPCA used in the TEP example. 
                                  (available from \\{https://github.com/Tveten/tpca})
  \item[R-package \texttt{tpcaMonitoring}:] Includes an implementation of Algorithm \ref{alg:tpcaMonitoring} and a single function to reproduce the entire simulation study in Section \ref{sec:5}.
                                            The package also contains .txt files with the results from our own run of the simulation study, to quickly recreate the figures.
                                            (available from {https://github.com/Tveten/tpcaMonitoring})
  \item[R-package \texttt{tdpcaTEP}:] All the code to easily reproduce the TEP results.
                                      (available from \\{https://github.com/Tveten/tdpcaTEP})
  % Each item, one file.
\end{description}

\renewcommand{\refname}{\large\sf REFERENCES}
\bibliography{library_natbib}
% \printbibliography

\newpage
\setcounter{section}{0}
\renewcommand{\thesection}{\Alph{section}}

\section*{Appendix A: Sensitivity of projections under other change distributions}
\begin{figure}[htb]
  \centering
  \includegraphics[scale = 0.7]{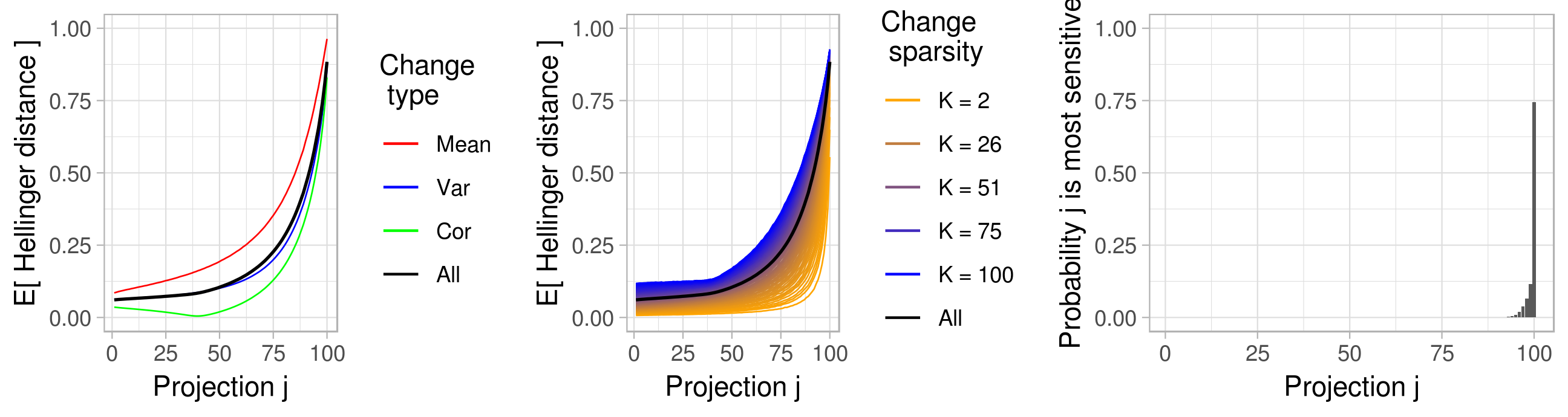}
  \caption{Monte Carlo estimates of $E[H_j]$ and $P_j$ with respect to the change distribution \eqref{eq:ex_change_distribution} and uniformly drawn pre-change covariance matrices $\bSigma_0$, with $D = 100$.
  (This is the same figures as shown in the main body of the article, for easier comparison with the figures below.)
  $10^3$ randomly drawn $\bSigma_0$'s were used, as well as $10^3$ Monte Carlo draws from the change distribution for each $\bSigma_0$}
  \label{fig:hellinger_summary_full_uniform_D100}
\end{figure}

Here we present a simulation study for investigating the robustness of our results to change distribution \eqref{eq:ex_change_distribution}, restated below for completeness.
\begin{equation}
\begin{split}
  \mathbf{C} &\sim \Multinom(p_\mu = 1/3, p_\sigma = 1/3, p_\rho = 1/3) \\
  K &\sim \Unif\{ 1, \ldots, D/2 \} \\
  \mu_d | K, \mathbf{C} &\iidsim \Unif[-1.5, 1.5],\; d \in \mathcal{D} \\
  \sigma_d | K, \mathbf{C} &\iidsim \frac{1}{2}\Unif[1/2.5, 1] + \frac{1}{2}\Unif[1, 2.5],\; d \in \mathcal{D} \\
  a_{di} | K, \mathbf{C} &\iidsim \Unif[0, 1),\; d \not= i \in \mathcal{D}.
\end{split}
\tag{\ref{eq:ex_change_distribution}}
\end{equation}
In the simulations, we set $D = 100$, and draw $10^3$ $\bSigma_0$'s uniformly from the space of correlation matrices \citep{joe_generating_2006}.
For each $\bSigma_0$, the sensitivity to changes is assessed as in Section \ref{sec:3}, by means of $10^3$ draws from a change distribution and calculating summary statistics of the Hellinger distances.
However, we now average the sensitivity results over the $10^3$ $\bSigma_0$'s, to get an average picture over many pre-change conditions.
Each of the figures presented below are therefore averages over $10^3$ figures like the ones shown in Section \ref{sec:3}
The average sensitivity results for change distribution \eqref{eq:ex_change_distribution} are shown in Figure \ref{fig:hellinger_summary_full_uniform_D100}.

There are four alternative change distributions we look at.
\begin{itemize}
  \item \textbf{Larger changes:} Mean interval $[-3, 3]$, standard deviation interval $[1/4, 4]$ (with the same split between decreases and increases) and correlation interval $[0, 0.5]$.
  \item \textbf{Smaller changes:} Mean interval $[-0.5, 0.5]$, standard deviation interval $[1/1.5, 1.5]$ and correlation interval $[0.5, 1]$.
  \item \textbf{Equal changes:} The same intervals as in \eqref{eq:ex_change_distribution}, but with all affected dimensions changing with the same value.
  For example, only one change size $\mu$ is drawn, and $\mu_d = \mu$ for all $d \in \mathcal{D}$.
\end{itemize}
The change parameters not mentioned have the same distribution as in \eqref{eq:ex_change_distribution}

\begin{figure}[htb]
  \centering
  \includegraphics[scale = 0.7]{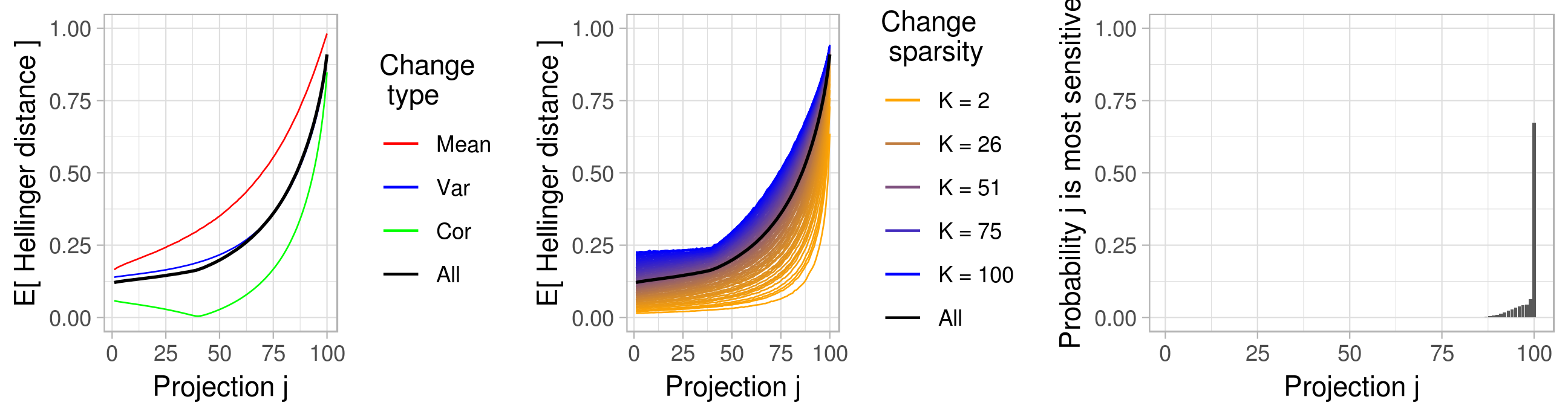}
  \caption{Monte Carlo estimates of $E[H_j]$ and $P_j$ with respect to the change distribution \eqref{eq:ex_change_distribution} and uniformly drawn $\bSigma_0$'s, but with \textbf{larger changes}}
  \label{fig:hellinger_summary_full_uniform_large_D100}
\end{figure}

\begin{figure}[htb]
  \centering
  \includegraphics[scale = 0.7]{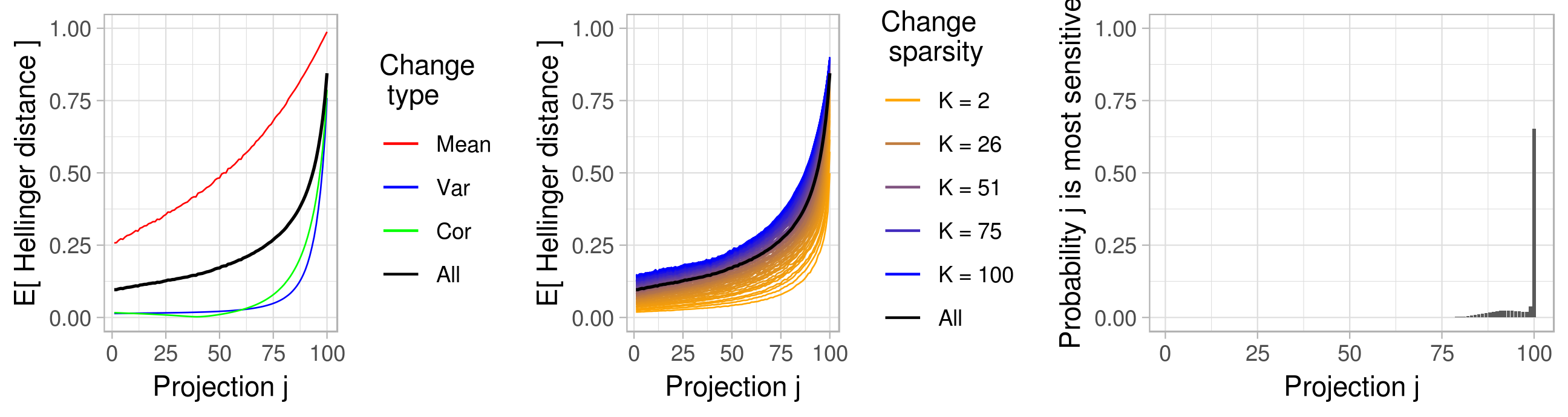}
  \caption{Monte Carlo estimates of $E[H_j]$ and $P_j$ with respect to the change distribution \eqref{eq:ex_change_distribution} and uniformly drawn $\bSigma_0$'s, but with \textbf{smaller changes}}
  \label{fig:hellinger_summary_full_uniform_small_D100}
\end{figure}

\begin{figure}[htb]
  \centering
  \includegraphics[scale = 0.7]{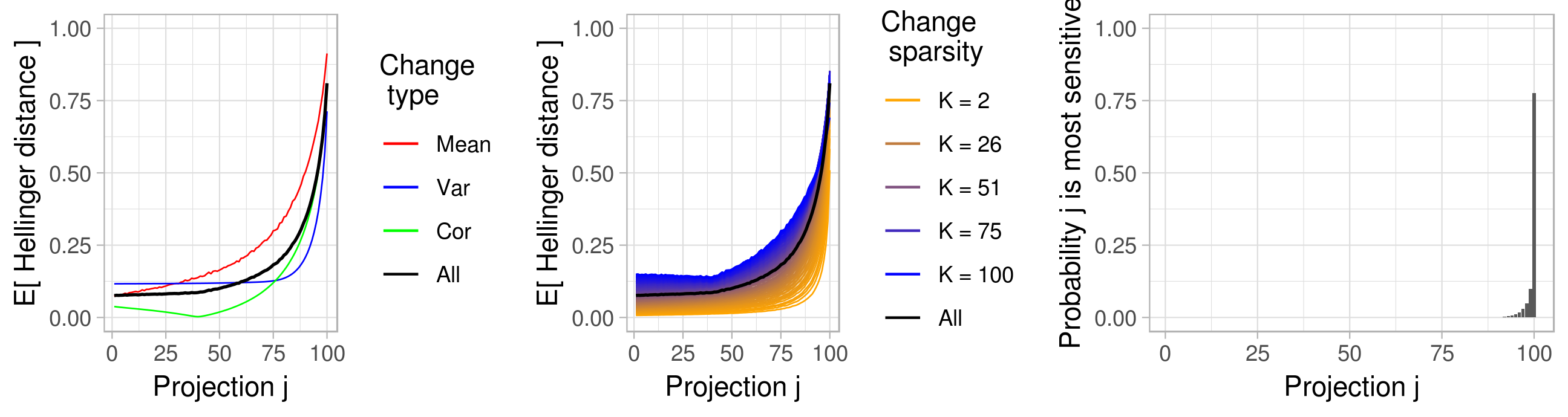}
  \caption{Monte Carlo estimates of $E[H_j]$ and $P_j$ with respect to the change distribution \eqref{eq:ex_change_distribution} and uniformly drawn $\bSigma_0$'s, but with a \textbf{equal changes} across all affected dimensions}
  \label{fig:hellinger_summary_full_uniform_equal_D100}
\end{figure}

Figures \ref{fig:hellinger_summary_full_uniform_large_D100} \ref{fig:hellinger_summary_full_uniform_small_D100} and \ref{fig:hellinger_summary_full_uniform_equal_D100} 
show the results for the larger, smaller and equal changes, respectively.
Overall, the sets of figures are very similar.
If one difference is to be mentioned, it is that for cutoff values $c$ close to $1$, both the larger and smaller changes would have resulted in slightly more projections being chosen by TPCA.

\begin{figure}[htb]
  \centering
  \includegraphics[scale = 0.7]{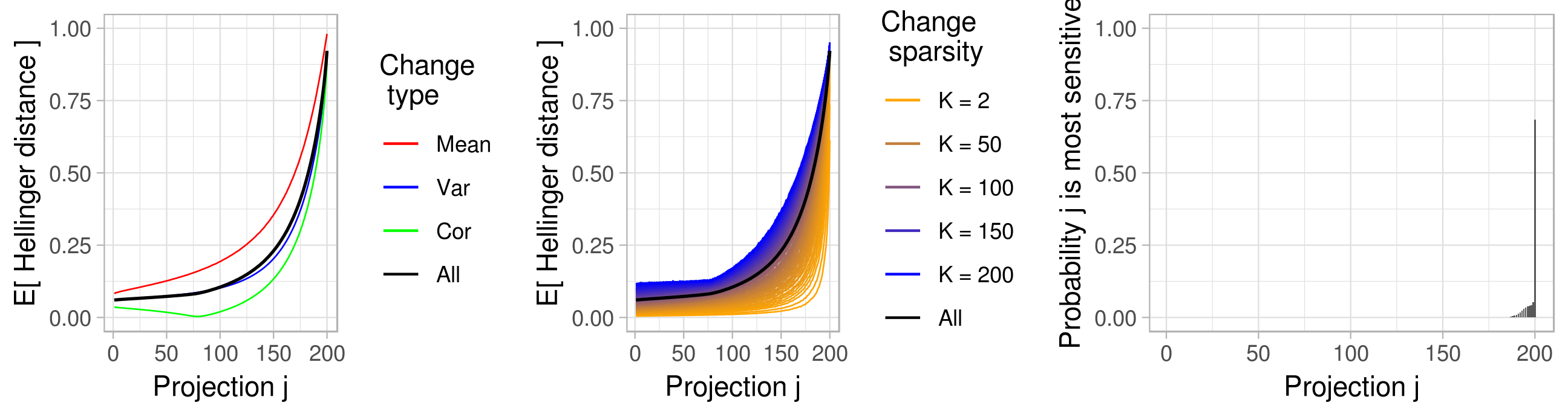}
  \caption{Monte Carlo estimates of $E[H_j]$ and $P_j$ with respect to the change distribution \eqref{eq:ex_change_distribution} and uniformly drawn $\bSigma_0$'s,
  but now with $D = 200$.}
  \label{fig:hellinger_summary_full_uniform_D200}
\end{figure}

Lastly, Figure \ref{fig:hellinger_summary_full_uniform_D200} shows the average sensitivity results with the same change distribution \eqref{eq:ex_change_distribution} but now with $D = 200$, 

\section*{Appendix B: Dealing with indefinite post-change correlation matrices}
When we change entries $\rho_{di}$ in the correlation matrix $\bSigma_0$ by multiplying them with factors $a_{di}$, it is not guaranteed that the changed matrix is positive definite.
To overcome this, we have used the function \texttt{nearPD} in the \texttt{Matrix} package of R.
This is a function that finds the nearest positive definite matrix to the input matrix in sup norm.
To obtain a correlation matrix, the diagonal is then put to $1$.

\end{document}